\documentclass[12pt,preprint]{aastex}
\usepackage[monochrome]{color}
\newcommand{\del}[2]%
{\frac{\mathrm{d}{#2}}{\mathrm{d}{#1}}}
\newcommand{\Del}[2]%
{\frac{\mathrm{D}{#2}}{\mathrm{D}{#1}}}
\newcommand{\ddel}[2]%
{\frac{\mathrm{d}^2{#2}}{\mathrm{d}{#1}^2}}

\newcommand{\pdel}[2]%
{\frac{\partial{#2}}{\partial{#1}}}
\newcommand{\pddel}[2]%
{\frac{\partial^2{#2}}{\partial{#1}^2}}
\newcommand{\udel}[1]{\partial_{#1}}
\newcommand{\tdel}[1]{\partial^{#1}}

\renewcommand{\vec}[1]{\mathbf{#1}}

\newcommand{\km}{\mathrm{km}}
\newcommand{\cm}{\mathrm{cm}}
\newcommand{\ms}{\ \mathrm{ms}}

\newcommand{\gauss}{\mathrm{G}}

\newcommand{\gpcmc}{\mathrm{g/cm^3}}
\newcommand{\MeV}{\ \mathrm{MeV}}
\newcommand{\ergps}{\ \mathrm{erg/s}}
\newcommand{\ergpcmc}{\ \mathrm{erg/cm^3}}

\def\Alfven{{Alfv\'{e}n}~} 
\def\d{\rm{d}}

\slugcomment{Accepted to the ApJ}
\shorttitle{SRMHD Simulations of Magnetically-dominated Jets}
\shortauthors{Takiwaki et al.}

\begin{document}

\title{Special Relativistic Simulations of Magnetically-dominated Jets
in Collapsing Massive Stars}

\author{Tomoya Takiwaki\altaffilmark{1},
Kei Kotake\altaffilmark{2,3},
 and Katsuhiko Sato\altaffilmark{1,4,5}}

\affil{\altaffilmark{1}Department of Physics,
School of Science, the University of Tokyo, 7-3-1 Hongo,
Bunkyo-ku,Tokyo 113-0033, Japan}
\email{takiwaki@utap.phys.s.u-tokyo.ac.jp}
\affil{$^2$Division of Theoretical Astronomy, National Astronomical Observatory of Japan, 2-21-1, Osawa, Mitaka, Tokyo, 181-8588, Japan}
\affil{$^3$Max-Planck-Institut f\"{u}r Astrophysik, Karl-Schwarzshild
-Str. 1, D-85741, Garching, Germany}
\email{kkotake@th.nao.ac.jp,kotake@MPA-Garching.MPG.DE}
\affil{\altaffilmark{4}
Research Center for the Early Universe,
School of Science, the University of Tokyo,7-3-1 Hongo,
Bunkyo-ku, Tokyo 113-0033, Japan}
\affil{\altaffilmark{5}
The Institute for the Physics and Mathematics of the Universe, The
University of Tokyo,
Kashiwa, Chiba, 277-8568 Japan}
\begin{abstract}
We perform a series of two-dimensional magnetohydrodynamic 
core-collapse simulations of rapidly rotating and strongly magnetized massive stars.
To study the properties of magnetic explosions for a longer time stretch 
of postbounce evolution,
we develop a new code under the framework of special relativity including 
 a realistic equation of state with a multiflavor neutrino leakage scheme.
Our results show the generation of the
magnetically-dominated jets in the two ways.
One is launched just after the core-bounce in a prompt way and another 
is launched at $ \sim 100 $ ms after the stall of the prompt shock.
We find that the shock-revival occurs when the 
magnetic pressure becomes strong, due to the field wrapping, enough
to overwhelm the ram pressure of the accreting matter.  The critical
toroidal magnetic fields for the magnetic shock-revival 
 are found to be universal of $\sim 10^{15}\mathrm{G}$ behind the jets.
 We point out that the time difference before the
shock-revival has a strong correlation with
 the explosions energies.
Our results suggest that the magnetically dominated jets are accompanied by the
formation of the magnetars. 
Since the jets are mildly relativistic, we speculate that they might be the origin of  
some observed X-ray flashes.
\end{abstract}

\keywords{supernovae: collapse, rotation ---
 magnetars: pulsars, magnetic field ---
 methods: numerical ---
 MHD --- special relativity ---
 gamma rays: bursts}

\section{Introduction}
There have been growing evidences shedding lights on the relations 
between the high energy astrophysical phenomena and their origins.
A number of host galaxies of long-duration gamma-ray bursts (GRBs) are 
recently identified as metal poor galaxies whose metalicities are 
lower than that of average massive star-forming galaxies \citep[]
[and reference therein]{sava06,stan06}.
The preponderance of short-lived massive star formation in such
young galaxies, as well as the identification of SN Ib/c light curves
peaking after the bursts in a few cases, has provided strong support for
a massive stellar collapse origin of the long GRBs \citep{pacz98,gala98,stan03}.
The duration of the long GRBs 
may correspond to the accretion of debris falling into the central black
hole (BH)\citep{piro98}, which suggests the observational consequence
of the BH formation likewise the supernova of neutron star formation.
There is also a growing observational evidence of supermagnetized
neutron stars with the magnetic fields of $\sim 10^{14}-10^{15}$ G, 
the so-called magnetars (\citet{dunc92}, see \citet{latt07} for
a recent review). 
The magnetic fields are determined by the measured period and derivative
of period under the assumption that the spin-down is caused due to the usual
magnetic dipole radiation \citep{zhan00,hard06}.
Tentative detections of spectral features during the burst phase also
indicate $B\sim 10^{15}\mathrm{G}$ when interpreted as proton cyclotron
lines \citep{gavr02,ibra03,rea03}.
 Recently X-ray flash (XRF), which is a low energy analogue of the GRB, is 
receiving great attentions as a possible
relevance to the magnetar formations \citep{maza06,toma07}.
A large amount of neutron rich Ni ejected by
SN2006aj associated with XRF060218 is interpreted to be the formation of such
objects, not the black hole after the explosion \citep{maed07a}.

So far a number of numerical simulations have been done towards the
understanding of the formation mechanisms of these compact
objects such as neutron stars, magnetars, and the black holes in
combination with their possible consequences like GRBs and XRFs. 
The leading model for the long-duration GRBs is the
collapsar model \citep{macf99}. In the model, the core of massive
stars with significant angular momentum collapses into a black hole (BH). 
The neutrinos emitted from the rotation-supported accretion disk
around the BH heat the matter of the funnel region of the disk, to
launch the GRB outflows. The relativistic flows are expected to ultimately form a
fireball, which is 
good for the explanation of the observed afterglow (e.g.,  \citet{pira99}).
In addition, it is suggested that the strong magnetic fields in the
cores of order of $10^{15} \gauss$ play also an active role both for driving 
the magneto-driven jets and for extracting a significant amount of
energy from the central engine (e.g.,
\cite{usov92,whee00,thom04,uzde07a} and see references therein). 

In order to understand such scenarios, 
the ultimate necessity of the stellar core-collapse simulations 
is to perform the simulations tracing all the phases
in a consistent manner starting from the stellar core-collapse, 
core-bounce, shock-stall, stellar explosion (phase 1) or BH formation
and the formation of accretion disk (phase 2), energy
deposition to the funnel region by neutrinos and/or magnetic fields
(phase 3), to the launching of the fireballs (phase 4). 
Here for convenience we call each stage as phase 1, 2, etc.
The requirement for the numerical modeling to this end is highly
computationally expensive, namely the multidimensional MHD
simulations not only with general relativity for handling the BH
formation, but also with the multi-angle neutrino transfer for
treating highly anisotropic neutrino radiation from the disks.
So various approximative approaches to each phase
have been undertaken.
As we mention below, these studies are complimentary in the sense that
the different epochs are focused on, with the different initial
conditions for the numerical modeling being taken.

In addition to the elaborate studies in the conventional supernova context 
(see recent reviews for \citet{kota06,jank07}), much attention has
been paid recently to the roles of rapid rotation and magnetic fields for studying 
the formation of magnetars and its possible application to
the collapsars {\color{red}
\citep{yama04,taki04,kota04a,sawa05,matt06,mois06,ober06a,nish06,burr07,cerd07,scei07,komi07b}.
}
After the failed or weak explosion,
the accretion to the central objects may lead to the formation of a BH (phase 2). 
Several general relativistic studies are on the line for the understanding 
of the hydrodynamics at the epoch of the BH formation, in which
more massive progenitors ($> \sim 25 M_{\odot}$) than those of the study
in the phase 1 are generally employed  \citep{shib06,seki07}. 
Treating the BH as an absorbing boundary or using the fixed metric
approaches, the numerical studies of 
the phase 3 are concerned with the initiation of the outflows from the funnel
region of the disk to the acceleration of the jets as a result of the
neutrino heating and/or MHD processes till the jets become mildly relativistic{\color{red}
\citep{koid98,macf99,prog03,nisi05,devi05,krol05,hawl06,mizu06,fuji06,uzde06,mcki07b,komi07a,naga07,suwa07a,suwa07b,bark08}.
}Numerical studies of the phase 4 are 
mainly concerned with the dynamics later on, namely, the jet
propagation to the breakout from the star, when the acceleration of
the jets to the high Lorentz factor is expected{\color{red}
\citep{ston00,aloy00,zhan03,leis05,mizt06,mcki06,mizu07}
}.

Our previous study was devoted to the phase 1, in which
 we performed a series of 2D core-collapse
simulations of rotating and magnetized massive stars under the framework of
the Newtonian magnetohydrodynamics \citep{taki04}. We found that the magneto-driven
jet-like shocks were launched from the protoneutron
stars just after core-bounce. However at the moment, we were unable to 
follow the dynamics much later on until when the collimated jets
reach further out from the center.
 The Alfv\'{e}n velocity of the jet
propagating into the outer layer of the iron core can be estimated by
 the following simple order-of-magnitude estimation,
\begin{equation}
v_{A} = \frac{B}{\sqrt{4 \pi \rho}} \sim 10^{10}\mathrm{cm/s}\frac{B/{10^{13}\mathrm{G}}}{\sqrt{\rho/\left(10^5\mathrm{g/cm^3}\right)}},
\end{equation}
with $\rho$ and $B$ being the typical density and magnetic field
 there. It can be readily inferred that the Alfv\'{e}n velocity 
can exceed the speed of light unphysically in the Newtonian simulation.
To avoid this problem we construct a new code under the framework of
special relativistic magnetohydrodynamics.
We take a wider parametric range for the strength of the rotation than
that of our previous work. By so doing, we hope to study
 more systematically than before how the strong magnetic
fields and the rapid rotation affect the properties of the magnetic 
explosions.

We summarize the numerical methods in section \ref{sec:NM}.
Section 3 is devoted to the initial models.
In section 4, we show the numerical results.
In section 5, we summarize our study and discuss the implications of
our model for the magnetars and the X-ray flashes.
Details of the numerical scheme and the code tests are given in the appendix.

\section{Numerical Methods}\label{sec:NM}
The results presented in this paper are calculated by the newly
developed special relativistic magnetohydrodynamic (SRMHD) code.
The novel point of this code is that the detailed microphysical
processes relevant for the stellar-core-collapse simulations
are also coupled to the magneto-hydrodynamics (MHD). 
We briefly summarize the numerical methods in the following.

The MHD part of the code is based on the formalism of \citet{devi03}.
Before going to the basic equations, we write down the definition of the
primary code variables.
The state of the relativistic fluid element at each point in the space time
is described by its density, $\rho$; specific energy, $e$; velocity,
$v^i$; and pressure, $p$.
And the magnetic field in the rest {\color{red}frame} of the fluid is described by
the 4-vector $\sqrt{4\pi}b^{\mu}={^*F}^{\mu\nu}U_{\nu}$, where $^*F^{\mu\nu}$ is the
dual of the electro-magnetic field strength tensor and $U_{\nu}$ is the
4-velocity.

After some mathematical procedures presented in Appendix \ref{app:DeBE},
the basic equations of SRMHD are described as follows:
\begin{eqnarray}
\pdel{t}{D}
+\frac{1}{\sqrt{\gamma}}\udel{i}{\sqrt{\gamma}Dv^i} &=&0 \label{eq:mass_consv}\\
\pdel{t}{E}
+\frac{1}{\sqrt{\gamma}}\udel{i}{\sqrt{\gamma}Ev^i}
&=&-p\pdel{t}{W}
-\frac{p}{\sqrt{\gamma}}\udel{i}{\sqrt{\gamma}W v^i} -  L_{\nu}\label{eq:ene_consv}\\
\pdel{t}{S_i-b^tb_i}
+\frac{1}{\sqrt{\gamma}}
\udel{j}{\sqrt{\gamma}\left(S_i v^j-b_ib^j\right)}
&=&
-\frac{1}{2}
\left(
\rho h \left(Wv_k\right)^2
- \left(b_k\right)^2
\right)\udel{i}{\gamma^{kk}}\nonumber\\
& &
-\left(\rho h W^2- {b^t}^2\right) \udel{i}{\Phi}\nonumber\\
& &
-\udel{i}{\left(p+\frac{\|b\|^2}{2}\right)}\label{eq:mom_consv}\\
\pdel{t}{B^i}
+\udel{j}{\left(Wv^jb^i-Wv^ib^j\right)}
&=&0\label{eq:induction}\\
\tdel{k}{{\udel{k}{\Phi}}}&=&\rho hW^2
-\left(p+\frac{\left|b\right|^2}{2}\right)
-\left(b^t\right)^2\label{eq:poisson}
\end{eqnarray}
where $W=\frac{1}{\sqrt{1-v^kv_k}}$, $D=\rho W$, $E=e W$ and 
$S_i=\rho hW^2v_i$ are 
 the Lorentz boost factor, auxiliary variables correspond to density,
energy, and momentum, respectively.
Eq. (\ref{eq:mass_consv}), Eq. (\ref{eq:ene_consv}) and Eq. (\ref{eq:mom_consv})
represents the mass, energy, and momentum conservations.
 $L_{\nu}$ in the right hand side of Eq. (\ref{eq:ene_consv})
is a total neutrino cooling rate determined by microphysical
processes which will be later explained.
In Eq. (\ref{eq:mom_consv}) it is noted that the relativistic enthalpy,
$h=(1+e/\rho+p/\rho+\left|b\right|^2/\rho)$
 includes magnetic energy.
Eq. (\ref{eq:induction}) is the induction equation for the magnetic
fields. $B^i$ are related to that in the rest frame of fluid 
as $B^i=Wb^i -Wb_t b^i$.
Here $b_t$ is a time component of the 4-vector, $b_{\mu}$.
Eq. (\ref{eq:poisson}) is the Poisson equation for the gravitational
potential, $\Phi$.

This newly developed code is an Eulerian code based on the finite-difference
method.
The numerical approach for solving the basic equations of
(\ref{eq:mass_consv}), (\ref{eq:ene_consv}), and 
(\ref{eq:mom_consv}), consists of the two steps, namely, 
the transport and the source step.
These procedures are essentially the same as those of ZEUS-2D \citep{ston92}.
At the transport step, the second order upwind scheme of Van Leer is implemented \citep{vlee77}.
To handle the numerical oscillations, we employ an artificial viscosity.
In the special relativistic treatments, many forms for the compression
heating are possible \citep{hawl84b}.
In our code, we employ the form of 
$\frac{\rho h}{\sqrt{\gamma}}\udel{i}{\sqrt{\gamma}Wv^{i}}$ as the
compression heating, which becomes the well-known artificial viscosity of
von Neumann and Richtmyer under the Newtonian approximation.
While not explicitly included in the above expression for the enthalpy, the
contribution from the compression heating on the inertia is included in our calculations.
The detailed status on the shock capturing using this term is shown at
Appendix \ref{TP}.

The time evolution of the magnetic fields is 
solved by induction equation, Eq. (\ref{eq:induction}).
In so doing, the code utilizes the so-called
constrained transport method, which ensures
the divergence free ($\nabla\cdot\vec{B}=0$) 
of the numerically evolved magnetic fields at all times.
Furthermore, the method of characteristics (MOC)
is implemented to propagate accurately all modes of MHD waves.
The detailed explanation and the numerical tests are delivered in the
appendix \ref{ap:AWP}.  
The self-gravity is managed by solving the Poisson equation, Eq. (\ref{eq:poisson})
with the incomplete Cholesky decomposition conjugate gradient method.

Together with these hydrodynamic procedures,
the following microphysical processes are implemented in this code. 
We approximate the neutrino transport by a multiflavor leakage
scheme \citep{epst81,ross03}, in which three neutrino flavors:
electron neutrino, $\nu_{e}$; electron antineutrino, $\bar{\nu}_{e}$; and
the heavy-lepton neutrinos, $\nu_{\mu}$, $\bar{\nu}_{\mu}$,
$\nu_{\tau}$, $\bar{\nu}_{\tau}$ (collectively referred to
as $\nu_{X}$), are taken into account. The 
neutrino reactions included are electron capture on
proton and free nuclei; positron capture on neutron; photo-, pair, plasma
processes \citep{full85,taka78,itoh89,itoh90}.
 We added a transport equation for the lepton fraction $Y_l (= Y_e -
Y_{e^{+}} + Y_{\nu_e} - Y_{\bar{\nu}_e})$,  
\begin{equation}
\pdel{t}{Y_l}
+\frac{1}{\sqrt{\gamma}}\udel{i}{\sqrt{\gamma}Y_l} = - \gamma_l
\end{equation}
to treat their 
change due to the relevant charged current reactions, whose reaction
rates are collectively represented by $\gamma_l$ here, with $Y_e,
Y_{e^{+}},Y_{\nu_e},Y_{\bar{\nu}_e}$, $\gamma_l$ being electron, positron,
electron neutrino, anti-electron neutrino fraction, respectively
(see \citet{epst81,ross03,kota03a} for details of the estimation of $\gamma_l$).
 $L_{\nu}$ in Eq. (\ref{eq:ene_consv}) represents the total 
neutrino cooling rate which is also estimated by the scheme. 
As for the equation of state (EOS), we employ a realistic one based on
the relativistic mean field theory \citep{shen98}.
Since the pressure is not represented as the analytic function of
density and internal energy like in the case of polytropic EOS,
an iterative algorithm are employed to update the fundamental variables 
(see Appendix \ref{CVFV} for detail).

In our two dimensional simulations, the spherical coordinate is
used with 300($r$) $\times$ 60($\theta$) grid points to cover the
computational domain. Axial symmetry and reflection symmetry across 
the equatorial plane are assumed.
The radial grid is nonuniform,
extending from $0$ to $4.0\times 10^8 \cm$ with finer
grid near the center. The finest grid is set to $10^5 \cm$.
The polar grid uniformly covers from $\theta=0$
to $\theta=\frac{\pi}{2}$. {\color{red} This choice of the grid numbers is
 sufficient for the aim of this paper as will be discussed in section 5.}

Finally we summarize the difference on the numerical approach from our
previous work \citep{taki04}.
Most major development is the fully special
relativistic treatment on magneto-hydrodynamics.
And for the microphysical parts  
the cooling terms by neutrino contains contributions from not only
$\nu_{\mathrm{e}}$ but also $\bar{\nu_{\mathrm{e}}}$ and
$\nu_{\mathrm{X}}$.
These advances provide more reliable results on the
magneto-rotational core-collapse.

\section{Initial Models}\label{sec:NGIM}

We make precollapse models by taking the profiles of density,
internal energy, and electron fraction distribution from
a rotating presupernova model of E25 by \citet{hege00}.
This model has mass of $25M_{\odot}$ at the zero age main
sequence, however loses the hydrogen envelope and becomes a Wolf Rayet
star of 5.45 $M_\odot$ before core-collapse.
Our computational domain involves the whole iron-core of $1.69 M_{\odot}$.
 It is noted that this model seems to be a good candidate as a progenitor of the GRB since
 the lack of the line spectra of the ejected envelopes are reconciled
with the observations of the supernovae associated with GRBs (e.g., \citet{mesz06}).

Since little is known about the spatial distributions of the rotation and
the magnetic fields in the evolved massive stars (see, however,
\citet{spru02}), we add the following 
rotation and magnetic field profiles 
in a parametric manner to the non-rotating core mentioned above.
For the rotation profile, we assume a cylindrical rotation of 
\begin{equation}
 \Omega(X,Z)=\Omega_{0}\frac{X_{0}^2}{X^2+X_{0}^2}\frac{Z_{0}^4}{Z^4+Z_{0}^4}\label{eq:CR},
\end{equation}
where $\Omega$ is the angular velocity and X and Z denote distance
from the rotational axis and the equatorial plane.
We adopt values of the parameters, $X_0$ and $Z_0$, as  $10^7 \cm,10^8
\cm$, respectively.
The parameter, $X_{0}$ represents the degree of differential rotation.
We assume the strong differential rotation as 
in our previous study \citep{taki04}.

As for the initial configuration of the magnetic fields, 
we assume that 
 the field is nearly uniform and parallel to the rotational axis in the core 
and dipolar outside.
For the purpose, we consider the following effective vector potential,
\begin{equation}
A_r=A_\theta=0,
\end{equation}
\begin{equation}
 A_\phi=\frac{B_0}{2}\frac{r_0^3}{r^3+r_0^3}r\sin\theta,\label{vec_phi}
\end{equation}
where $A_{r,\theta,\phi}$ is the vector potential in the $r,\theta,\phi$ 
direction, respectively,  $r$ is the radius, $r_0$ is the radius of the core, and 
$B_0$ is the model constant.
In this study, we adopt the value of $r_0$ as $2\times10^8$ cm which
is approximately the size of the iron core at a precollapse stage.
This vector potential can produce the uniform magnetic fields when $r$ 
is small compared with $r_0$, and the dipole magnetic fields for vice versa.
Since the outer boundary is superposed at $r=4\times10^8~{\rm cm}$, the magnetic
fields are almost uniform in the computational domain 
as the previous work \citep{taki04}.
It is noted that this is a far better way than the loop current method 
for constructing the dipole magnetic fields \citep{symb84}, 
because our method produces no divergence of the magnetic fields 
near the loop current. 
We set the outflow boundary conditions for the magnetic fields at the 
outer boundary of the calculated regions.


We compute 9 models changing
the total angular momentum and the strength of magnetic fields
by varying the value of $\Omega_0$ and $B_{0}$.
The model parameters are shown in Table \ref{tab:model}.
The models are named 
after this combination,
with the first letters, B12, B11, B10,
representing strength of the initial magnetic field,
the following letter, TW4.0, TW1.0, TW0.25 
representing the initial $T/|W|$, respectively.
Here $T/|W|$ indicates the ratio of 
the rotational energy to the absolute value of the gravitational energy.
The corresponding values of $\Omega_{0}$ are 
$151 \mathrm{rad/s}$, $76 \mathrm{rad/s}$, $38 \mathrm{rad/s}$ 
for TW4.0, TW1.0, TW0.25, respectively.
It is noted that the value of $T/|W|$ is $0.15\%$ of the progenitor by 
\citet{hege00} and also that the specific angular momenta ranges from 
$0.5$ to $1.5$ $j_{16}$ for TW0.25 to TW4.0 models 
with $j_{16}\equiv 10^{16}~\rm{cm}^2~{\rm s}^{-1}$, 
which are in good agreement with the requirement of the collapsar model \citep{macf99}.
{\color{red}
Current stellar evolution calculations predict that the rapidly rotating 
massive stars with smaller metalicity experiences the so-called chemically homogeneous
 cores during its evolution \citep{yoon}. 
Such stars are considered to satisfy the requirements of the collapsar model, 
namely rapid rotation of the core \citep{woos06}.
According to a GRB progenitor model of 35OB in \citet{woos06}, 
the magnetic field strength of the core reaches to $\sim 10^{12}$ G and 
the specific angular momentum is the order of $j_{16} \sim 1$), with which our choices 
for the initial magnetic field and the initial rotation rate are reconciled.}

\begin{table}[h]
\begin{center}
\caption{Models and Parameters }\label{tab:model}
\begin{tabular}{cc|ccc}
\tableline\tableline
 & &        & {$T/|W|(\%)$}& \\
 & & 0.25\% &  1.0\%       &4.0\%\\
\tableline
 &$10^{10}\gauss$  & B10TW0.25 & B10TW1.0  & B10TW4.0 \\
$B_{0}(\mathrm{Gauss})$ & $10^{11}\gauss$  & B11TW0.25 & B10TW1.0  & B11TW4.0 \\
 &$10^{12}\gauss$  & B12TW0.25 & B10TW1.0  & B12TW4.0 \\
\tableline
\end{tabular}
\tablecomments{
Model names are labeled by the initial strength of magnetic fields and rotation.
$T/|W|$ represents the ratio of the rotational energy to the absolute value of
  the gravitational energy.
The corresponding values of $\Omega_{0}$ in  Eq. (\ref{eq:CR}) are 
$151 \mathrm{rad/s}$, $76 \mathrm{rad/s}$, $38 \mathrm{rad/s}$ 
for TW4.0, TW1.0, TW0.25, respectively.
$B_{0}$ represents the strength of the poloidal magnetic fields (see Eq.(\ref{vec_phi})).
The corresponding values of $E_{m}/|W|$ is $2.5\times 10^{-8}$,
 $2.5\times 10^{-6}$ and $2.5\times 10^{-4}$ for $10^{10}\gauss$,
 $10^{11}\gauss$ and $10^{12}\gauss$, respectively with $E_{m}$ being
the magnetic energy.
}
\end{center}
\end{table}

\clearpage

\section{Results}
\subsection{Hydrodynamics before Core-Bounce}
First of all, we briefly mention the dynamics before core bounce,
when the gross features are rather similar among the computed models.
The characteristic properties are summarized in Table \ref{tab:PROCPS}.

The story before core-bounce is almost the same as the canonical 
core-collapse supernovae with rapid rotation (see, e.g.,
\citet{kota03a}). The core begins to collapse 
due to electron captures and the photodissociation of the iron nuclei,
and eventually experiences the bounce at the subnuclear density 
 by the additional support of the centrifugal forces.
In fact, the central densities at bounce becomes smaller and the epoch
till bounce is delayed as the initial
rotation rates become larger (see $\rho_{\rm bnc}$ and $t_{\rm bnc}$ 
in Table \ref{tab:PROCPS}).

\begin{figure}[ht]
 \begin{center}
\includegraphics[width=.99\linewidth]{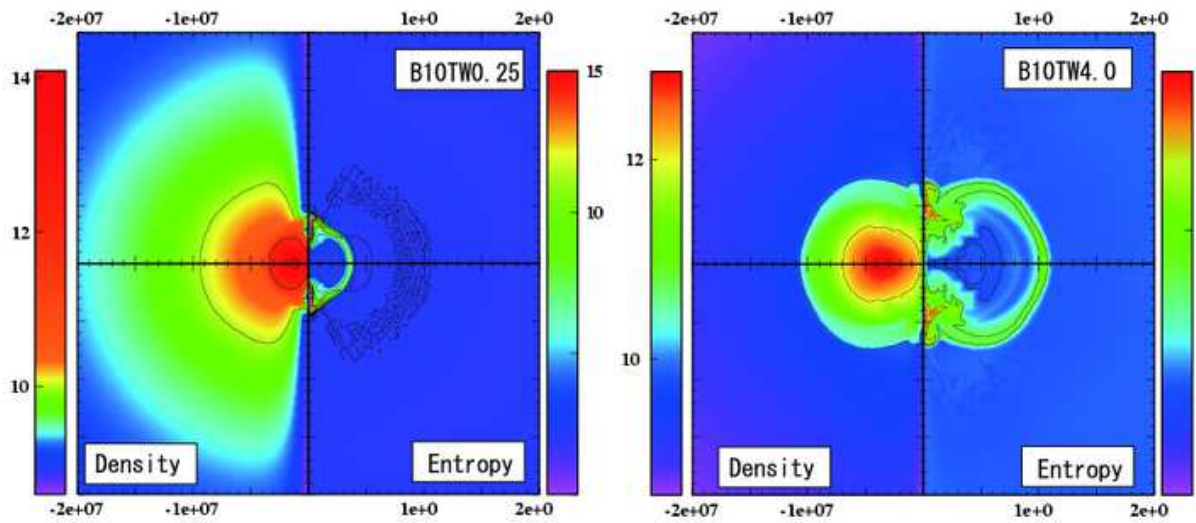}
\caption{Snapshots for models B10TW0.25 (left) and
B10TW4.0 (right), showing the degree of the rotational flattening at core bounce.
In each panel, contour of density [$\mathrm{g/cm^3}$] (left) and entropy per
  baryon [$k_B$] (right) is shown. Flattening of the unshocked core 
(the central low entropy region seen in the right side of each panel) is
remarkable for the right panel.
Note that the unit of the horizontal and the vertical axis is in cm.
}\label{fig:ds_bnc}
 \end{center}
\end{figure}

As the compression proceeds,
the rotational energy increases and reaches a few $10^{52}
\mathrm{erg}$ at the moment of the bounce 
(seen from $T/|W|_{\rm bnc} \times W_{\rm bnc}$ in Table
\ref{tab:PROCPS}). 
Given the same initial rotation rates, the values of $T/|W|_{\rm bnc}$ 
 do not depend on the initial field strength so much. This means that the 
angular momentum transfer is negligible before bounce, which is also
the case of the Newtonian hydrodynamics \citep{yama04}.
At bounce, the unshocked core becomes more flattened 
as the initial rotation rate becomes larger (compare panels in Figure \ref{fig:ds_bnc}).
\textcolor{red}{The central protoneutron stars rotate very rapidly reaching to $\sim 
3000$ rad/s with the typical surface magnetic fields of $\sim 10^{13}$ 
G to $\sim 10^{15}$ G for B10 and B12 models, respectively.}
From the table, it is also seen that the amplification rates of the magnetic
fields ($A_{\mathrm{amp}}$) 
are mainly determined by the initial rotational rates.
One exception is the model B12TW4.0. Due to very rapid rotation with the highest
magnetic fields initially imposed, the model bounces predominantly due
to 
the magnetic force. As a result, the core bounce occurs 
earlier with the lower central density with less gravitational
energy of the inner core than the models with the same initial rotation rate 
(see Table \ref{tab:PROCPS}). This earlier magnetically-supported bounce 
 leads to the suppression of the amplification rate, which is exceptionally
observed for this model.

 In this way, the hydrodynamic properties before bounce are
 mainly governed by the differences of the initial rotation rates. 
On the other hand, the differences of the magnetic field strength begin to play an important role on
the dynamics later on.
We will mention them in detail from the next sections.

\begin{table}[h]
\caption{MHD properties till core bounce}\label{tab:PROCPS}
\begin{tabular}{cccccccccc}
\tableline\tableline
Model&
 $T/|W|_{{\mathrm{bnc}}}$  &
 $\rho_{\mathrm{bnc}}$  & $|W_{\mathrm{bnc}}|$  &
 $E_m/|W|_{\mathrm{ini}}$ & 
 $E_m/|W|_{\mathrm{bnc}}$ &
 $\frac{E_{p}}{E_m}$ & 
 $A_{\mathrm{amp}}$ & $t_{\mathrm{bnc}}$ \\
 Names &  & {\small $[10^{14}\mathrm{g/cm^3}]$ }  & $[10^{53}\mathrm{erg}]$  &
  & & & & $[\mathrm{ms}]$ \\
\tableline
B12TW0.25 &
 $0.10$ &
 $2.1$ & $1.1$ &
 $2.5 \times 10^{-4}$ & $1.0 \times 10^{-3}$ &
 $0.3$ &
 $100$ & $245$ \\
B11TW0.25 & 
 $0.10$ &
 $2.1$ & $1.1$ &
 $2.5 \times 10^{-6}$ & $1.0  \times 10^{-5}$ &
$0.3$ &
 $100$ &  $245$ \\
B10TW0.25 &
 $0.10$ &
 $2.1$ & $1.1$ &
 $2.5 \times 10^{-8}$ & $1.0  \times 10^{-7}$ &  
$0.3$ &
 $100$ & $245$ \\
\tableline
B12TW1.0 &
 $0.18$ &
 $1.3$ & $1.1$ &
 $2.5 \times 10^{-4}$ & $9.0 \times 10^{-3}$ &
$0.07$ &
  $720$ & $295$ \\
B11TW1.0 & 
 $0.18$ &
 $1.3$ & $1.1$ &
 $2.5 \times 10^{-6}$ & $7.0  \times 10^{-5}$ & 
$0.07$ &
 $610$ & $295$ \\
B10TW1.0 &
 $0.18$ &
 $1.3$ & $1.1$ &
 $2.5 \times 10^{-8}$ & $7.0  \times 10^{-7}$ & 
$0.07$ &
$610$ &  $295$ \\
\tableline
B12TW4.0 &
 $0.20$ &
 $0.095$ & $0.68$ &
 $2.5 \times 10^{-4}$ & $20 \times 10^{-3}$ & 
$0.3$ &
 $800$ & $477$ \\
B11TW4.0 & 
 $0.19$ &
 $0.11$ & $0.74$ &
 $2.5 \times 10^{-6}$ & $29  \times 10^{-5}$ & 
$0.1$ &
$4400$ &  $484$ \\
B10TW4.0 &
 $0.19$ &
 $0.11$ & $0.74$ &
 $2.5 \times 10^{-8}$ & $31  \times 10^{-7}$ & 
$0.1$ &
$4400$ &  $484$ \\
\tableline
\end{tabular}
\tablecomments{
Characteristic properties before core bounce.
$T/|W|_{\mathrm{bnc}}$ is the rotational energy per gravitational energy
 at bounce. 
 $\rho_{\mathrm{bnc}}$ is the maximum density at bounce. 
 $E_m/|W||_{\mathrm{ini}}$ and $E_m/|W||_{\mathrm{bnc}}$
 is the magnetic energy per the gravitational energy initially and at
bounce, respectively.
 $\frac{E_{p}}{E_m}$ is the ratio of the poloidal magnetic energy to the total
 magnetic energy at bounce.
$A_{\mathrm{amp}}$ represents the amplification rate of magnetic
energy until core bounce, which is defined as $A_{\mathrm{amp}}
\stackrel{\mathrm{def}}{\equiv} (E_m|_{\mathrm{bnc}})/(E_m|_{\mathrm{ini}})$.
$t_{\mathrm{bnc}}$ represents the time till bounce.
}

\end{table}

\clearpage

\subsection{Prompt vs. Delayed MHD Exploding Model}\label{ssec:pvd}
\begin{figure}[ht]
 \begin{center}
\includegraphics[width=.66\linewidth]{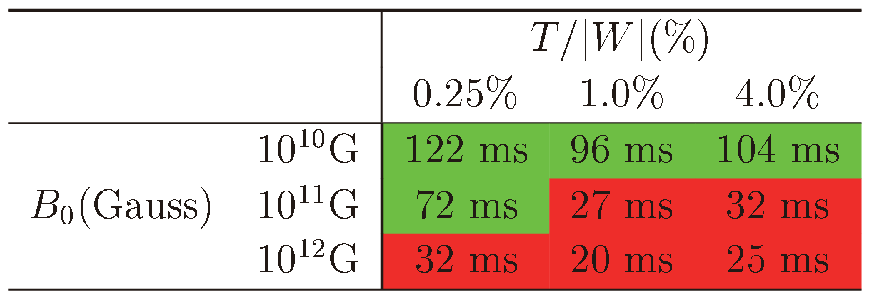}
\caption{Classification of the computed models into the prompt (red
blocks) or delayed (green blocks) MHD exploding model by the difference 
of $t_{1000\mathrm{km}}$ shown in this table, which is the 
shock-arrival time to the radius of $1000~{\rm km}$ after bounce.
}\label{fig:two_groups}
 \end{center}
\end{figure}

After bounce,  
we can categorize the computed models into two groups, by the criterion whether
the shock generated at bounce promptly reach the surface of the
 iron core or not.
For later convenience, we call the former and the latter models as 
prompt and delayed MHD exploding model, respectively throughout the
paper.
The models and the corresponding groups are shown in Figure \ref{fig:two_groups}.
To begin with, we choose typical model from the two groups and mention
their properties in detail.

\begin{figure}[h]
 \begin{center}
  \includegraphics[width=.99\linewidth]{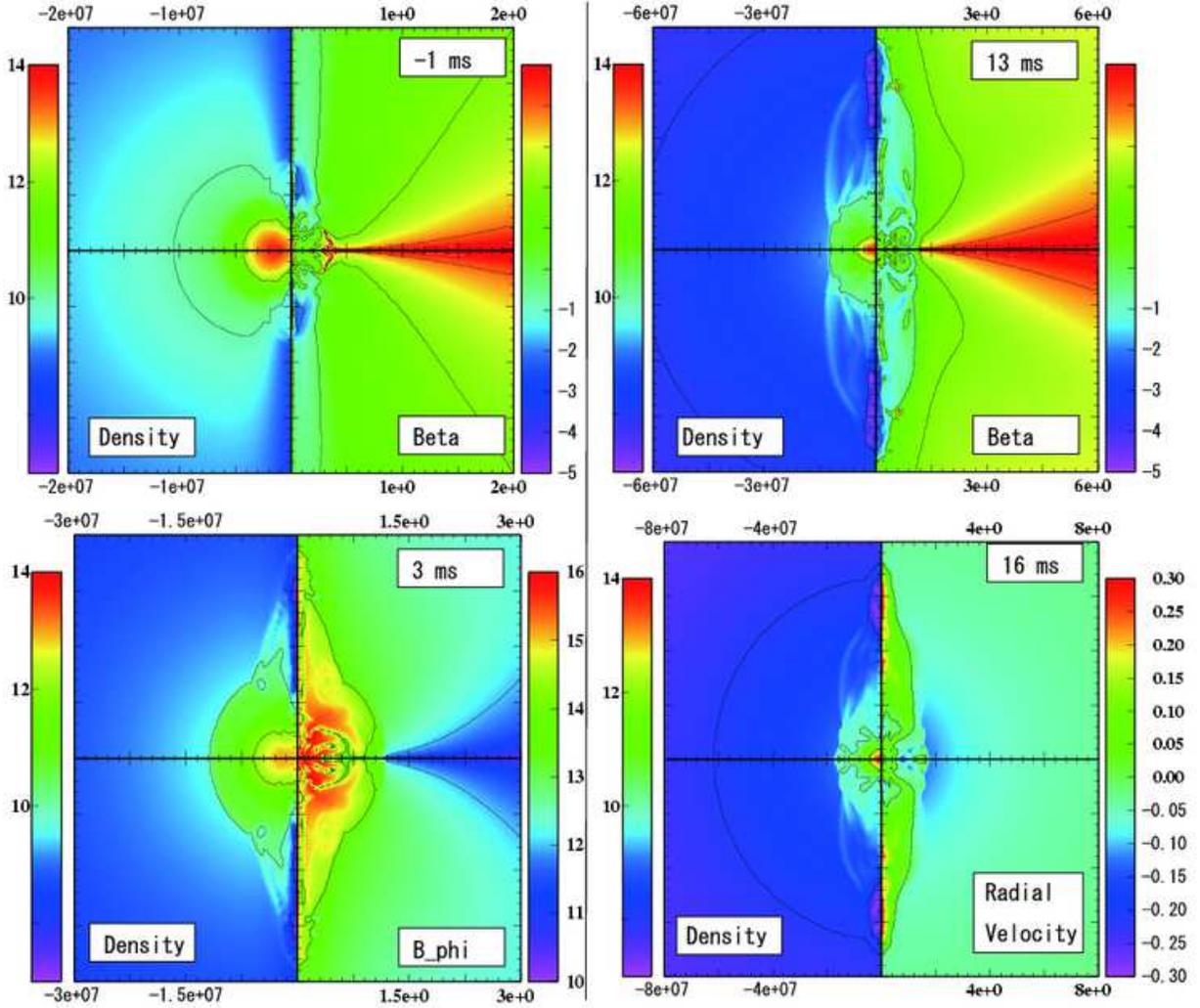}
\caption{Time evolution of various quantities characterizing 
the dynamics near after bounce 
for a prompt MHD exploding model 
(see text in section 4.2 of Prompt MHD exploding models). 
This is for model B12TW1.0.
In each panel, the left side
represents the logarithm of density [$\mathrm{g/cm}^3$]. Time in each
panel is measured from the epoch of bounce. At the top panels, the right
side is {\color{red} the logarithm of} the ``plasma $\beta$'' $\stackrel{\mathrm{def}}{\equiv} p/\frac{B^2}{8\pi}$, indicated by ``Beta''. At the bottom left
panel, the right side is the logarithm of toroidal component of the magnetic fields
  [$\mathrm{G}$], indicated by ``$B_{\rm phi}$''. 
At the bottom right panel, the right side is the radial
velocity in unit of the speed of light $:c$.
Note that the unit of the horizontal and the vertical axis of all
panels are in cm.
}\label{fig:b12tw100seq}
 \end{center}
\end{figure}

\paragraph{Prompt MHD Exploding Model}
The models classified into this group have strong magnetic
fields and rapid rotation initially.
Figure \ref{fig:b12tw100seq} shows the dynamics near core bounce. As
seen from the bottom left panel, the shock at core bounce stalls
 in the direction to the equatorial plane
 at $\sim 1.4 \times 10^{7} \cm$ promptly ($\sim 3$ ms) after bounce. However the shock in the direction of the
 rotational axis does not stall and becomes a collimated jet (see top
right and bottom right). The wound-up magnetic fields are an important
agent to explain these properties.
 
The magnetic fields for the promptly MHD models are {\color{red}strong enough} to power
the jet already at the epoch of bounce.
That is clearly shown in the top left panel, showing that the ``plasma 
$\beta$'' $\stackrel{\mathrm{def}}{\equiv} p/\frac{B^2}{8\pi}$, being the ratio of {\color{red}the matter to
the magnetic pressure}, outside the unshocked core near
the poles becomes very low (typically $10^{-2}$). From the right side
of the bottom left panel, the toroidal magnetic field strength there
reaches over $10^{15}\mathrm{G}$. 
The dynamics around the poles are strongly affected by these strong magnetic fields.

\begin{figure}[h]
 \begin{center}
  \includegraphics[width=.99\linewidth]{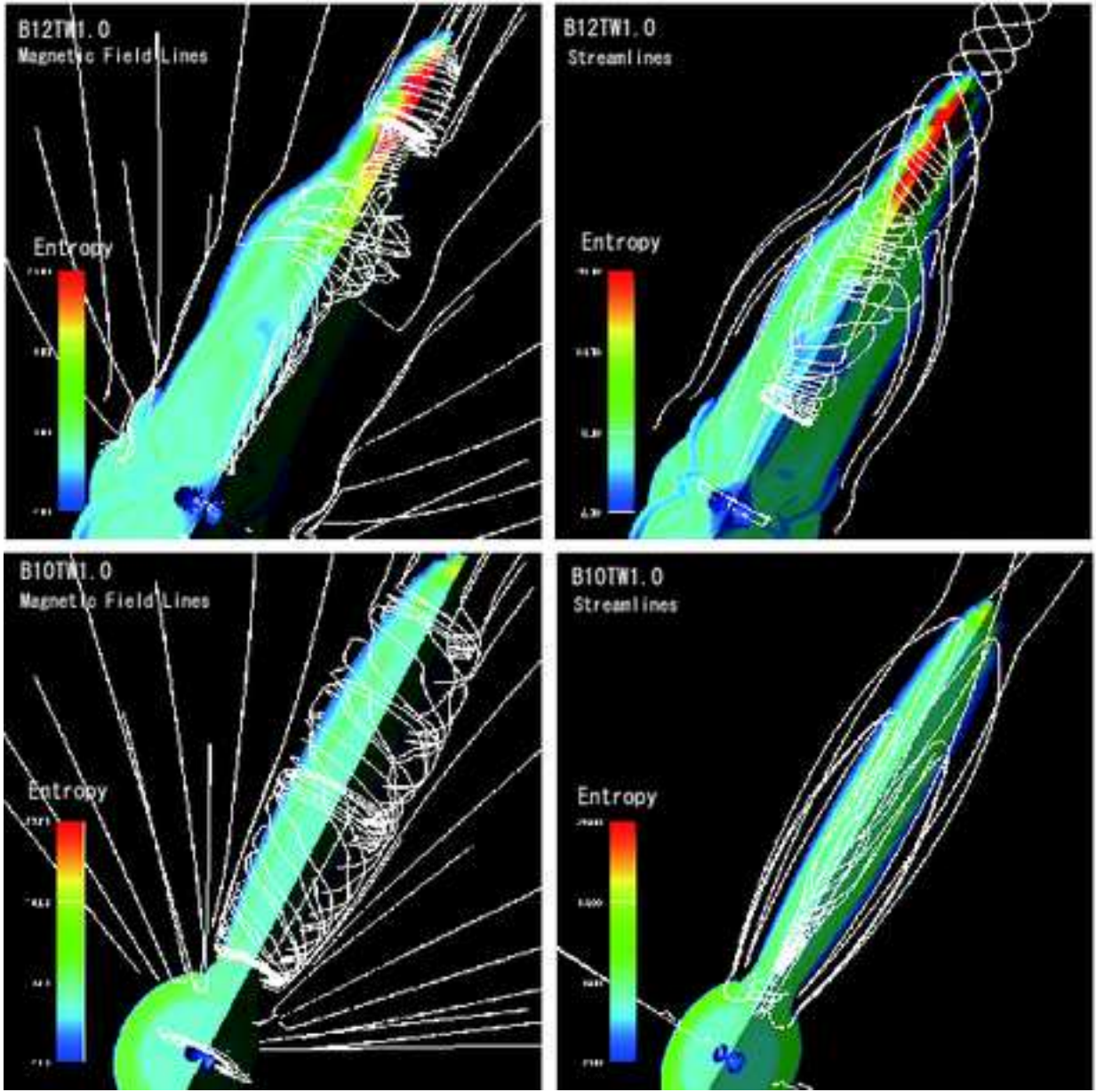}
\caption{
Three dimensional plots of entropy with the magnetic field lines
(left) and the streamlines of the matter (right) during the jet
propagation for models of B12TW1.0 (top) and B10TW1.0 (bottom), at 
$20 \ms$ and $94 \ms$ after bounce, respectively.
The outer edge of the sphere colored by blue represents the radius of 
$7.5 \times 10^7 \cm$.
 Note that the model of the top and bottom panel belongs to 
the prompt and delayed MHD exploding model, respectively.
These panels highlight not only the wound up magnetic field around the
rotational axis (left), but also the fallback of the matter from the head of the
jet downwards to the equator, making a cocoon-like structure behind
the jet (right).
}\label{fig:3djet}
 \end{center}
\end{figure}

The three dimensional plots of Figure \ref{fig:3djet} are 
useful to see how the field wrapping occurs. 
From the top left panel, it is seen that the field lines are strongly
wound around the rotational axis.
The white lines in the top right shows 
the streamlines of the matter.
A fallback of the matter just outside of the head of the jet downwards
to the equator (like a
cocoon) is seen. In this jet with a cocoon-like structure, 
the magnetic pressure is always {\color{red} dominant over} the matter pressure (see
the region where plasma $\beta$ less than 1 in the right side of the
top right panel of Figure \ref{fig:b12tw100seq}).
\textcolor{red}{
This magneto-driven jet does not stall and penetrate to the surface of
the iron core, which is essentially the reproduction of the pioneering 
results in the MHD supernova simulations by \citet{leblanc} and its analysis by 
 \citet{meier}.}
The speed of the head of the jet is mildly relativistic of $\sim 0.3 c$,
with $c$ being the speed of light (the right side of bottom right panel of Figure \ref{fig:b12tw100seq}).
At $20 \ms$ after bounce, the jet finally reaches the surface of the
iron core of $\sim 10^8 \cm$.
At this moment, the explosion energy, which will be a useful quantity for comparing
the strength of the explosion among the models later,
 reaches $1.4 \times 10^{50}\  \mathrm{erg}$.

\clearpage

\paragraph{Delayed MHD Exploding Model}
The models with weaker initial magnetic fields belong to the delayed 
MHD exploding model (see Figure \ref{fig:two_groups}).
In the following, we explain their properties taking model B10TW1.0 as an example.
It is noted again that this model has the same initial 
rotation rate with model B12TW1.0 of the previous section, 
but with the two orders-of-magnitudes weaker initial magnetic fields.

\begin{figure}[h]
 \begin{center}
  \includegraphics[width=.8\linewidth]{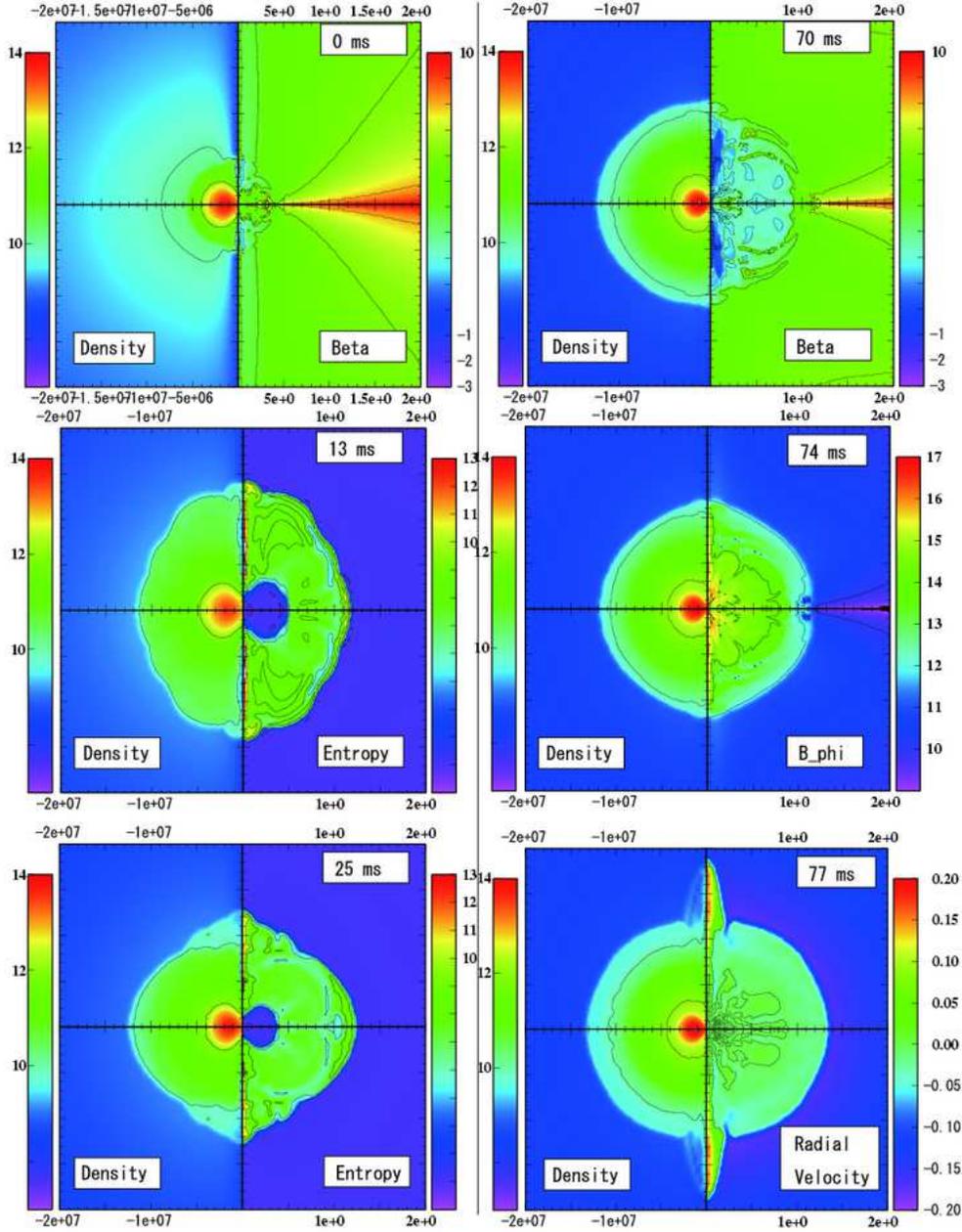}
\caption{Same as Figure 3 but for the quantities showing the dynamics near 
after bounce for a delayed exploding model (see text in section 4.2:Delayed MHD 
exploding model). This is for model B10TW1.0. During the oscillations
of the stalled shock after bounce (from middle left to
bottom left), the magnetic fields behind the stalled shocks become
large enough, due to the field wrapping (top and middle right), leading to the
shock-revival for the formation of the magnetically-dominated jet (bottom right).
Note that the unit of the horizontal and the vertical axis of all
panels is in cm.
}\label{fig:b10tw100seq}
 \end{center}
\end{figure}

In the case of model B10TW1.0, the shock wave at bounce stalls 
in all directions at $\sim 1.5 \times 10^{7} \cm$.
As shown in the top left panel of Figure
\ref{fig:b10tw100seq}, the plasma $\beta$ is so high that the magnetic
fields play no important role before bounce.
After the shock stalls, the stalled shock begins to oscillate.
The middle left and the bottom left panel shows the prolate and 
oblate phase during the oscillations, respectively.
Until $\sim 70 \ms$ after bounce, the oscillation of the shock front
continues diminishing its amplitude. Approximately the number of the
oscillations is about $5$ times this time. Without the magnetic
fields, the oscillation should cease settling into the equilibrium
state with the constant accretion through the stalled-shock to the center. However during this
oscillation, the magnetic fields behind the stalled shock 
gradually grow due to the field wrapping and the plasma $\beta$ around the
polar regions becomes low as seen from the right side of the top right panel.
Soon after the toroidal magnetic fields become as high as $\sim 10^{15}\mathrm{G}$
behind the stalled shock (see the middle right panel),
the stalled shock near the pole suddenly begins to propagate along the
rotational axis and
turns to be a collimated jet (see the bottom right panel).
This revived jet does not stall in the iron-core. This is the reason
why we call this model as the delayed MHD exploding model.
The speed of the jet reaches about $5.5\times 10^9 \mathrm{cm/s}$ (see
the bottom right panel).
Also in this jet, the toroidal component of the magnetic fields is
dominant over the poloidal one and a fallback of the matter
is found in the outer region of the jet (cocoon) as in the case of the
promptly MHD exploding model (see the bottom two panels of Figure \ref{fig:3djet}).
At $\sim 96 \ms$ after bounce, the jet reaches $\sim 10^8 \cm$.
The explosion energy at that time reaches
$0.094 \times 10^{50} \mathrm{ergs}$.

As mentioned, the dynamical behaviors between the prompt and delayed MHD
exploding models after bounce seem apparently different. However
 there are some important similarities between them, which we discuss 
from the next section. 
 
\clearpage
\subsection{Similarities of Prompt and Delayed MHD Exploding Model}

In this section we focus on the similarities between the prompt and
delayed MHD exploding models.

\begin{figure}[ht]
 \begin{center}
 \includegraphics[width=.90\linewidth]{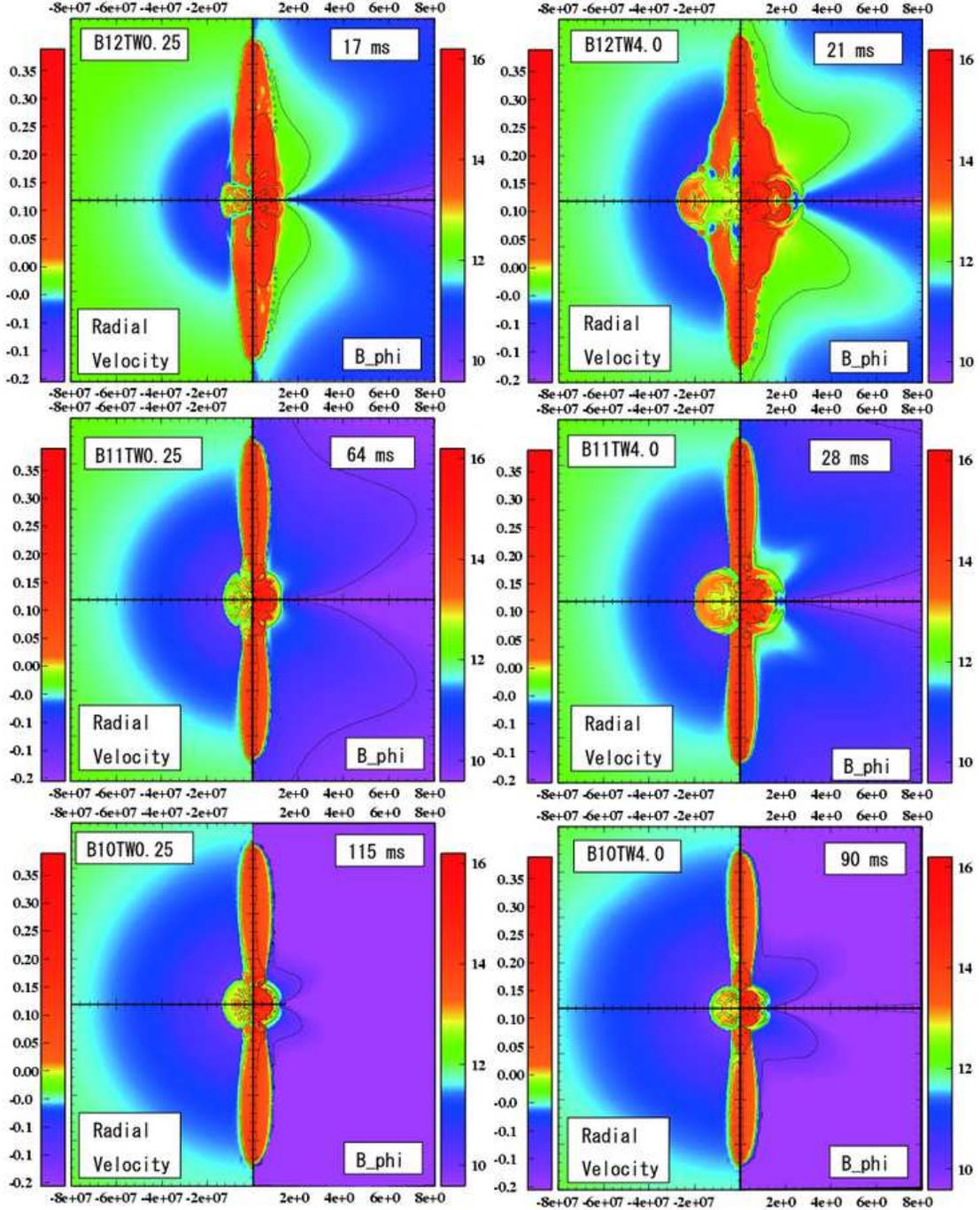}
\caption{Contour of radial velocity (left side) and toroidal magnetic
fields (right side in each panel) for various models, showing the similarity of the jets between the
prompt and delayed MHD exploding models.
The right and left panels correspond to the rapidly rotating ($T/|W|=4.0\%$)
and the slowly rotating $T/|W|=0.25\%$ models, respectively.
From top to bottom panels, the initial strength of the
magnetic fields changes from strong (B12 models) to weak (B10 models).
The time from bounce is shown in the top right part of each panel,
indicating the difference between the prompt and the delayed models.  
Note that the unit of the horizontal and the vertical axis of all
  panels is in cm.}\label{fig:vrB}
 \end{center}
\end{figure}

From Figure \ref{fig:vrB}, it can be seen that 
the radial velocities and the magnetic fields of the jets are quite
similar among the models regardless of the prompt or delayed exploding
models.
Typical values of the toroidal magnetic fields are $10^{14}-10^{16}\mathrm{G}$
and typical velocities are $10-30\%$ of the speed of light.
The opening angles of the jets are also similar.
The width of this jet is about $8 \times 10^6 \cm$ when the jet
reaches $7.5\times 10^7 \cm$, which means that the half opening angle
of the jets is about $6^{\circ}$ at this time.
These characteristic values of the jets are summarized in Table \ref{tab:PROJET}.

\begin{figure}[ht]
 \begin{center}
  \includegraphics[width=.99\linewidth]{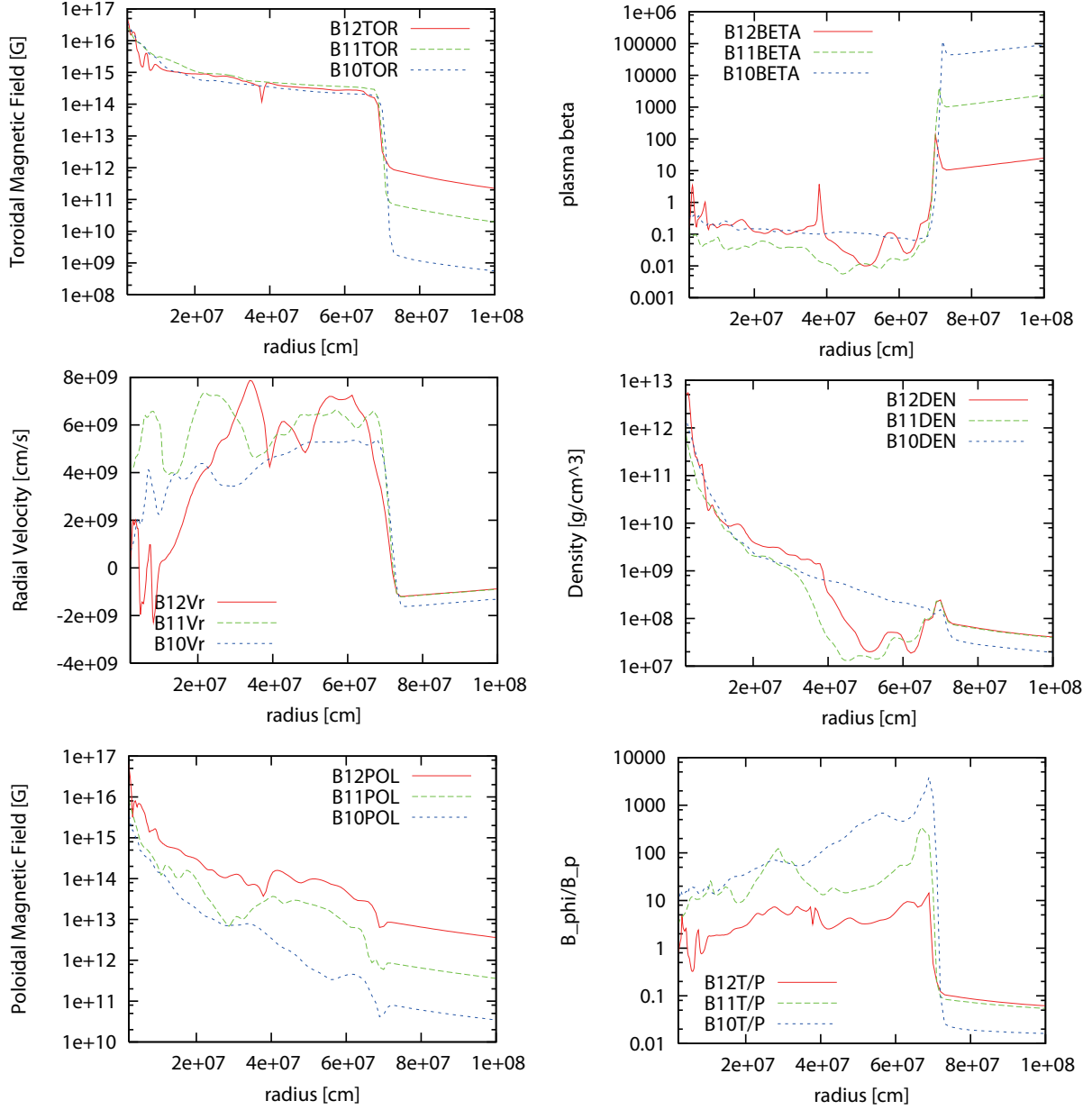}
\caption{
Various profiles of jets along the rotational axis, for models
 B12TW1.0 at $20 \ms$ (solid), B11TW1.0 at $24 \ms$ (short-dashed)
 and B10TW1.0 (dotted) at $94 \ms$ after bounce, respectively, showing
the effects of the difference of the initial magnetic fields (fixing
the initial rotation rate of $T/|W|=1.0\%$).
Top left and right is the toroidal magnetic field ($B_{\phi}$) and
plasma beta ($\beta = P/\frac{B^2}{8\pi}$).
Middle left and right is the radial velocity [cm/s] and the density [$\mathrm{g/cm^3}$].
Bottom left and right is the poloidal magnetic field ($B_{p}$) and the ratio
of toroidal to poloidal magnetic field ($B_{\phi}/B_{p}$). 
Note that the shock position is approximately $700$ km as seen from
the discontinuity of these profiles.
{\color{red}
It should be noted that the toroidal fields here are not  
 just at $\theta=0^{\circ}$ (the fields are zero there) 
but at the closest mesh to the axis of 
$\theta=1.5^{\circ}$.}
}\label{fig:JPTW100}
 \end{center}
\end{figure}

\begin{figure}[ht]
 \begin{center}
   \includegraphics[width=.99\linewidth]{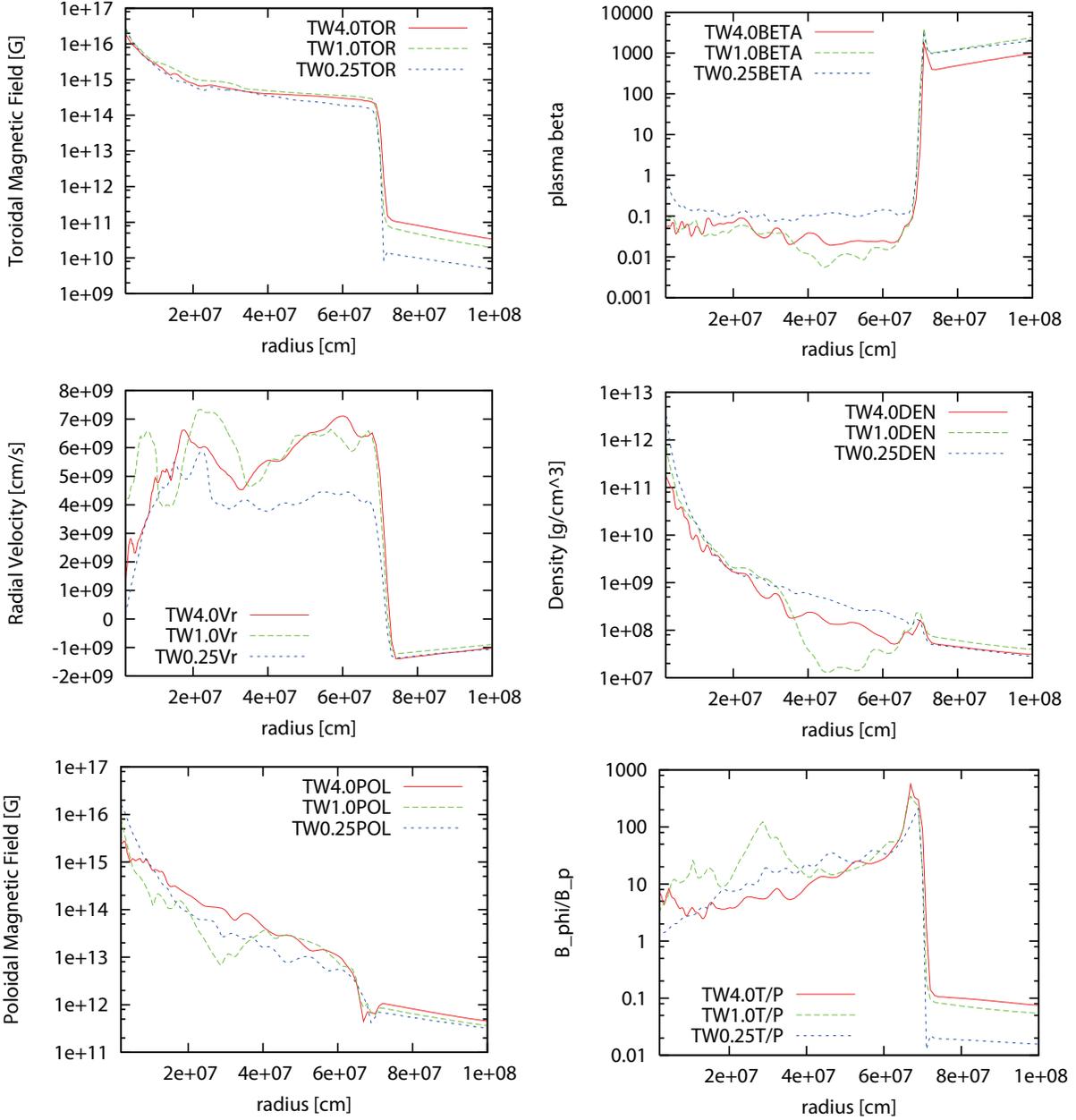}
\caption{Same as Figure \ref{fig:JPTW100}, but for models 
B11TW4.0 at $28 \ms$ (solid), B11TW1.0 at $24 \ms$ (short-dashed),
 and B11TW0.25 (dotted) at $64 \ms$ after bounce, respectively,
showing the effects of the difference of the initial rotation rates 
(fixing the initial field strength of $B11$).
}\label{fig:JPB11}
 \end{center}
\end{figure}

Detailed properties of the jets {\color{red} in the vicinity along the rotational axis}
 are shown in Figures \ref{fig:JPTW100}
and \ref{fig:JPB11} to see the origin of these similarities.
 We fix the initial rotation
rate and the initial field strength in Figure \ref{fig:JPTW100} and
\ref{fig:JPB11}, respectively, to see their effects separately.
In Figure \ref{fig:JPTW100}, the initial rotation rate is
$T/|W|=1.0\%$ and the different lines correspond to the
difference between the initial magnetic fields from $10^{12}$ (B12) to
$10^{10}$ G (B10). In Figure \ref{fig:JPB11}, the initial magnetic
field is $10^{11}\mathrm{G}$ and the different lines corresponds the
difference in the initial rotation rates. 

From the top and middle panels of Figures \ref{fig:JPTW100} and
\ref{fig:JPB11}, we find that the radial profiles of the 
toroidal magnetic field,
the plasma $\beta$ ($0.1-0.01$), the density,
and the velocity, are rather similar behind the shock whose position 
can be seen from the discontinuity at $\sim 700$ km.
Above all, it is surprising to see the
remarkable similarity in the profiles of the toroidal magnetic fields
behind the shock among the models (top left in Figures \ref{fig:JPTW100} and
\ref{fig:JPB11}). The typical strength behind the shock 
is seen to be $\sim 10^{15}\mathrm{G}$. This critical strength of the toroidal
magnetic field for the shock-revival is estimated as follows. 
The matter behind the stalled-shock is pushed inwards by the ram pressure of
 the accreting matter. This ram pressure is estimated as,
\begin{equation}
 P=4 \times
  10^{28}\left(\frac{\rho}{10^{10}\mathrm{g/cm^3}}\right)\left(\frac{\Delta
  v}{2 \times 10^9\mathrm{cm/s}}\right)^2\mathrm{erg/cm^3}, 
\end{equation}
where the typical density and the radial velocity are taken 
from Figure \ref{fig:ds_bnc} and the bottom right panel of 
Figure \ref{fig:b10tw100seq}, respectively.
When the toroidal magnetic fields are amplified as large as $\sim
10^{15}\mathrm{G}$ due to the field wrapping behind the shock,
the resulting magnetic pressure, $\frac{B^2}{8\pi}$, can overwhelm the ram
pressure, leading to the magnetic shock-revival. 
The origin of the similarity of the jets seen in
Figure \ref{fig:b12tw100seq} comes from this mechanism.
We find that this process works in all the computed models.
{\color{red}It is noted that the importance of the magnetic-shock revival 
was noticed also in the analytic models by \citet{uzde07a,uzde07b}.
In addition to their expectations, our simulations show that the 
explosion energy becomes smaller than their estimations because 
the magnetic tower cannot be wider as they assumed.}

From the bottom panels of Figure \ref{fig:JPB11}, it can be seen
 that the poloidal fields behind the shock front 
do not depend on the initial rotation rate so much given the same
initial field strength,
while the difference of the poloidal magnetic fields behind the shock 
in the bottom panels of Figure \ref{fig:JPTW100} simply comes from the
difference in the initial field strength. 
This feature is regardless of the prompt or delayed models.

\clearpage
\begin{table}
 \begin{center}
\caption{Characteristic properties of jets }\label{tab:PROJET}
\begin{tabular}{cccccc}
\tableline\tableline
Name&
 $t_{\mathrm{exp}}$ & $t_{\mathrm{1000km}}$ &
 $v_{\mathrm{jet}}$ & ${E_{\mathrm{exp}}}_{\mathrm{1000km}}$ \\
  & $[\mathrm{ms}]$ & $[\mathrm{ms}]$  &
  $[\mathrm{cm/s}]$ & $[10^{50}\mathrm{ergs}]$ \\
\tableline
B12TW0.25 &
 $ 0$ & $ 32 $ & $ 6.0 \times 10^{9}$ & $ 1.3$ \\
B11TW0.25 &  
 $48$ & $ 72 $ & $ 3.7 \times 10^{9}$ & $0.05$ \\
B10TW0.25 &
 $92$ & $ 122$ & $ 3.8 \times 10^{9}$ & $0.02$ \\
\tableline
B12TW1.0 &
 $ 0$ & $ 20 $ & $ 8.0 \times 10^{9}$ & $1.4$   \\
B11TW1.0 & 
 $10$ & $ 27 $ & $ 8.0 \times 10^{9}$ & $0.23$  \\
B10TW1.0 &
 $70$ & $ 96 $ & $ 5.5 \times 10^{9}$ & $0.094$ \\
\tableline
B12TW4.0 &
 $ 0$ &  $ 25$ & $ 6.0 \times 10^{9}$ & $1.0$   \\
B11TW4.0 & 
 $16$ &  $ 32$ & $ 6.0 \times 10^{9}$ & $0.10$  \\
B10TW4.0 &
 $65$ &  $104$ & $ 5.5 \times 10^{9}$ & $0.006$ \\
\tableline
\end{tabular}
\tablecomments{
Properties of the jets.
$t_{\mathrm{exp}}$ is the duration from bounce to the revival of the
stalled shock due to the field wrapping. 
$t_{\mathrm{1000km}}$ represents the duration required for the jet to
 reach $\sim \mathrm{1000km}$ after bounce.
Explosion energy, ${E_{\mathrm{exp}}}_{\mathrm{1000km}}$, and the jet
  velocity, $ $ $v_{\mathrm{jet}}$ is estimated at the moment.
${E_{\mathrm{exp}}}_{\mathrm{1000km}}$ are normalized as $10^{50}$erg.
For the definition of the explosion energy, see Eq. (\ref{eq:expene}).
}
\end{center}
\end{table}
\clearpage

\subsection{Dependence of Jet Arrival Times and Explosion Energies on 
Initial Rotation Rates and Magnetic Field Strengths}\label{sec:dependence}
In the previous section, we discuss the similarity among the computed models.
From this section, we move on to discuss the differences among them.

\paragraph{Jet Arrival Time}
First we discuss the ``jet arrival time'' shown in Table
\ref{fig:two_groups}, which is the timescale when the jet reaches the outer edge of
the iron core of $\sim 1000\mathrm{km}$.
As discussed in the previous section, this timescale 
is mainly determined how long it takes for the magnetic fields behind the
shock to become as large as the critical toroidal magnetic fields 
($\sim 10^{15}\mathrm{G}$) as a result of the field wrapping.

From the top left panel of Figure \ref{fig:exptime}, 
it is seen that the strong initial magnetic fields shorten the jet arrival time.
This tendency is seen in all the computed models regardless of the
prompt or delayed exploding models.
When the initial magnetic fields are strong enough ($\sim 10^{-4}$ of
the gravitational energy), the jet arrival times between the different
initial rotational models become almost the same. 
 In this case, the critical magnetic fields for
the shock-revival are already generated by the
compression before core bounce. So the strong magneto-driven jets
 can produce the prompt MHD explosions in a similar way.
For the rapidly rotating
models (the sequence of TW1.0 and TW4.0), 
it is seen that the decrease in the rate of the jet arrival time
as a function of the initial $E_m/|W|$ becomes smaller 
when the initial $E_m/|W|$ is larger than $\sim 10^{-6}$ (see the kink
in the panel). This is because too strong
magnetic fields transport the angular momentum of the protoneutron
star outwards, leading to the suppression of the efficiency of the
field wrapping after bounce.

In the top right panel of Figure \ref{fig:exptime},  the 
 dependence of the jet arrival time on the initial rotation rate is shown. 
By intuition, the jet arrival time may become shorter as the initial
rotation rates become larger since the field-wrapping should become
more efficient. The panel shows that this is true for moderately
rotating models of the initial $T/|W|$ less than $0.01$, 
but not true for the more rapidly rotating models. This can be
 explained as follows.
Too rapid rotation of the core hinders the central core from collapsing 
due to the stronger centrifugal forces.
This feature is clearly shown in the middle left panel of Figure
\ref{fig:exptime} showing the density profiles.
The density near the center is $\sim 100$ times lower than that for the slowly
rotating models. 
{{\color{red} 
Since the angular momentum is well conserved before bounce (see section 4.1),
 the inner core ($\lesssim 20$ km) gains smaller angular velocities 
for rapidly rotating models by the weakened compression 
as seen in the middle right panel of Figure \ref{fig:exptime}.}
Reflecting these aspects, the amplification rate
of the magnetic fields ($\del{t}{E_m}/E_m$) near core-bounce becomes smaller 
for the most rapidly rotating model (TW4.0) as seen from the bottom panel of Figure 9.
This suppression makes the jet arrival time almost constant or longer as the initial
$T/|W|$ becomes larger than $\sim 0.01$ as in the top right panel of
Figure \ref{fig:exptime}.

\paragraph{Explosion Energies}
In addition to a wide variety of the jet arrival times,
we find a large difference in the strengths of the magnetic
explosions.

As a measure of the strength, we define the explosion energy as,
\begin{equation}
{E_{\mathrm{exp}}}_{\mathrm{1000km}}
=
\int_{\mathrm{D}}\d
V~e_{\mathrm{local}}
=
\int_{\mathrm{D}}\d
V\left(e_{\mathrm{kin}}+e_{\mathrm{int}}+e_{\mathrm{mag}}+e_{\mathrm{grav}}\right),\label{eq:expene}
\end{equation}
here $e_{\mathrm{local}}$ is the sum of $e_{\mathrm{kin}}$, $e_{\mathrm{int}}$, $e_{\mathrm{mag}}$ and
$e_{\mathrm{grav}}$, with being the kinetic, internal, magnetic, and gravitational energy, respectively (see Appendix \ref{app:enedef} for
their definitions in special relativity) and $\mathrm{D}$ represents
the domain where the local energy is positive, indicating that the matter
is not bound by the gravity. The explosion energy is evaluated when
the jet arrives at the radius of $1000\km$ at the polar direction. 
The value of the explosion energy is summarized in Figure
\ref{fig:expenergy}.
Generally speaking, it is found that the explosion energies becomes
larger for the prompt MHD exploding models (red) than the delayed MHD
exploding models (green).

What makes the difference on the explosion energies 
{\color{red} 
at the shock breakout from the iron cores?}
{\color{red} Firstly, the initial strength of the magnetic field 
is the primary agent to affect the explosion energies. The explosion energies are 
larger for models with the larger initial fields as seen in Figure 11.
Secondly, the geometry 
of the jets has also effects on the explosion energies.}
Figure \ref{fig:ble} shows the toroidal magnetic fields (left side)
and the local energy (right side) in the jets from the
stronger to the weak magnetic fields models (from top to bottom panels)
{\color{red}  at the shock breakout}. {\color{red} 
In each right panel, it is noted 
that the regions with the positive local energies 
($e_{\mathrm{local}} > 0$ in Eq. (12)) are drawn with color scales and 
the regions with black are for the regions with the negative local energies.}
It is seen that the regions where 
the local energy is positive mostly coincide with the regions where 
the strong toroidal magnetic fields are generated.
As the initial field strength becomes larger, the regions where the
local energy becomes positive, becomes larger ({\color{red} i.e. the jets become wider}),  leading to the larger explosion energies. 
In the case of the delayed
exploding model (the right panel in Figure \ref{fig:ble}), 
it is found that the width of the jets becomes
narrower, which results in the smaller explosion energies.
Although the properties of the jets just on the rotational axis are
similar among the models seen from Figures 7 and 8,
 the lateral structures of the
jets is found to have influence over the explosion energies.


\begin{figure}[h]
 \begin{center}
   \includegraphics[width=.99\linewidth]{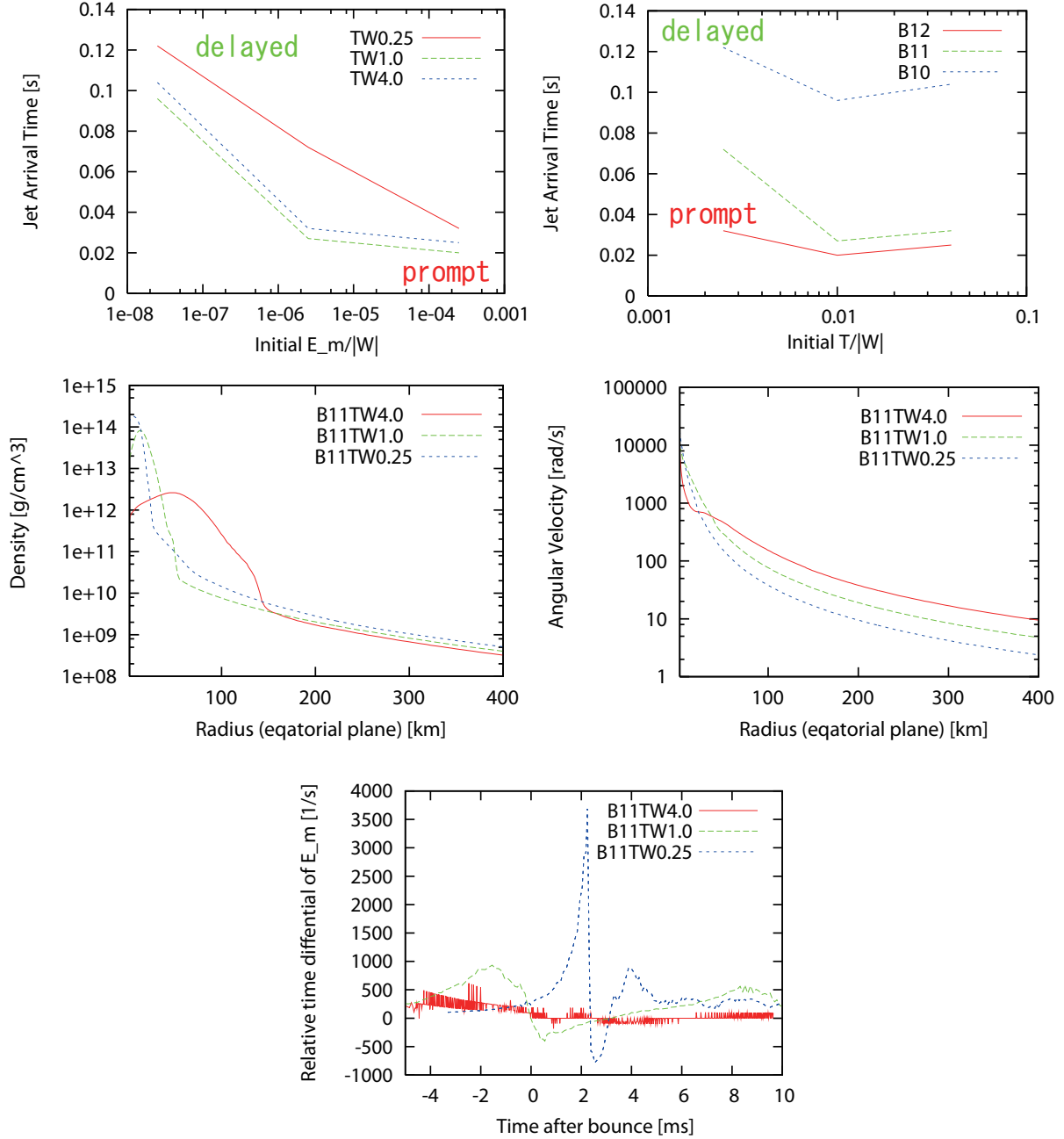}
\caption{Dependence of the jet arrival time on 
the initial magnetic field strength (top left) and the initial rotation rate 
(top right). Here
$E_m/|W|$ and $T/|W|$ represent the ratio of initial magnetic and
rotational energies to the gravitational energy, respectively. The jet arrival time
is the duration for the jet to reach the outer edge of iron core of
$\sim 1000\mathrm{km}$. Middle left and right panel shows the distribution
of the density and the angular velocity for models B11TW4.0, B11TW1.0 and
  B11TW0.25 at core bounce as a function of the equatorial radius, respectively.
Bottom panel shows the temporal amplification rate ($\del{t}{E_m}/E_m$) near
core bounce.
}\label{fig:exptime}
 \end{center}
\end{figure}
\clearpage

\begin{figure}[ht]
 \begin{center}
  \includegraphics[width=.66\linewidth]{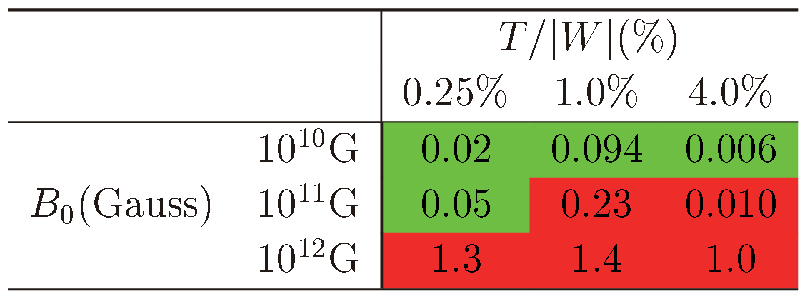}
\caption{
The explosion energy for all the models.
The energies are normalized to $10^{50} \mathrm{erg}$.
Note again that the red and green blocks indicate the prompt and
delayed MHD exploding models, respectively. 
See Eq. (\ref{eq:expene}) for the definition of the explosion energy.
}\label{fig:expenergy}
 \end{center}
\end{figure}
\clearpage

\begin{figure}[h]
 \begin{center}
  \includegraphics[width=.99\linewidth]{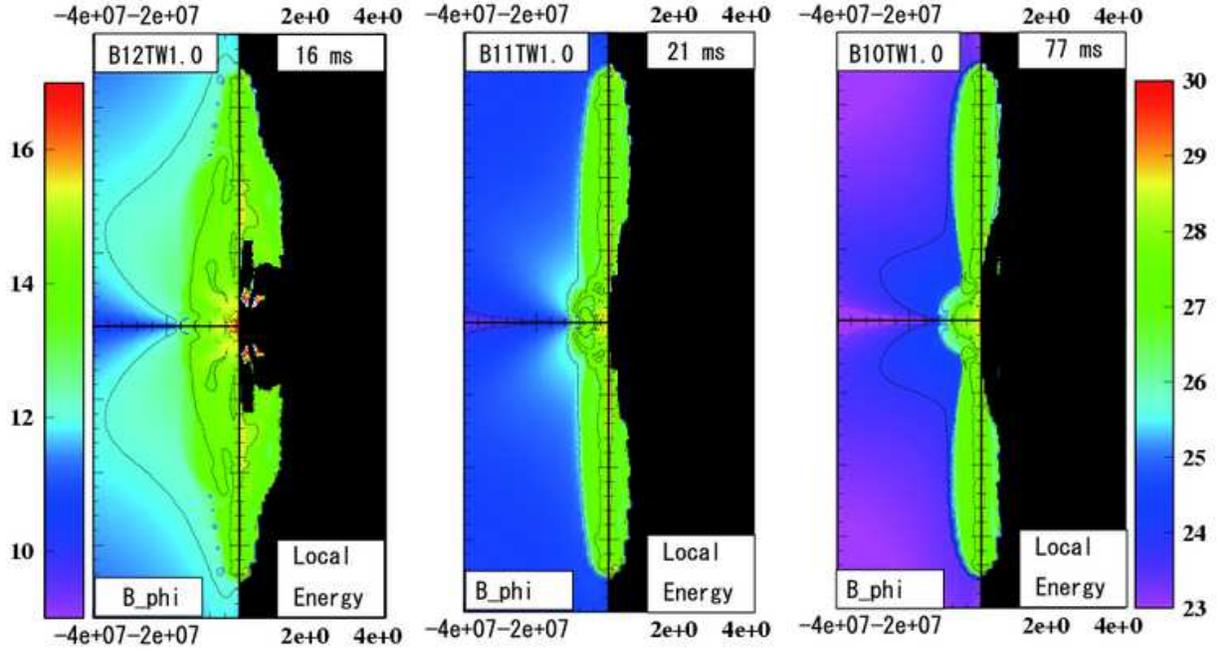}
\caption{Contour of toroidal magnetic field and the 
local energy for models B12TW1.0, B11TW1.0 and B10TW1.0 near the shock breakout 
from the iron core.
In each panels, the logarithm of toroidal magnetic field[G] (left
side) and the logarithm of local energy [$\mathrm{ergs/cm}^3$] (right
side) is shown, respectively. It is noted for each right panel 
that the regions with the positive local energies ($e_{\mathrm{local}} > 0$ in Eq. (12))
are drawn with color scales and 
the regions with black are for the regions with the negative local energies.
The timescales given is measured from the core-bounce indicate the prompt or delayed MHD 
exploding models. The unit of the horizontal and the vertical axis of all
  panels are in cm.
}\label{fig:ble}
 \end{center}
\end{figure}
\clearpage

\section{Summary and Discussion}
We performed a series of two-dimensional MHD
simulations of rotational core-collapse of magnetized massive stars.
The main motivation was to clarify how the strong magnetic fields and 
the rapid rotation of the core affect the magnetic explosions.
To handle the very strong magnetic fields, 
we developed a new code under the framework of special relativity.
A novel point is that the microphysics such as the realistic equation
of state and the neutrino cooling are implemented to the special relativistic MHD code.
Due to these advantages, our computation can achieve 
a longer time-stretch of the evolutions compared to previous studies.
The obtained results can be summarized as follows.
\begin{itemize}
\item 
Magnetically powered jets are commonly found in all the
computed models. In the jets, the magnetic fields are dominated by the
toroidal components as a result of the field wrapping. For the
profiles and strengths of the toroidal fields behind the jets, we find a remarkable
similarity. We find that 
the jet-like explosions occur when the 
magnetic pressure behind the shock becomes strong, due to the field
wrapping, enough to overwhelm the ram pressure of the accreting matter.  The required
toroidal magnetic fields are similar of $\sim 10^{15}\mathrm{G}$, 
which can be also understood by a simple order-of-magnetite estimation.
Reflecting the similarity in the mechanism of producing jets, 
global properties of the jets such as the velocities ($\sim 20\%$ of
the speed of light) and the half opening
angle of the jets ($\sim 6^{\circ}$) are also found to be similar among the computed
models.


\item The timescale before the onset of the magnetic shock-revival are 
quite different depending on the initial strengths of rotation and 
 magnetic fields. When the initial strengths of rotation and
magnetic field are larger, the jet can be launched just after the
core-bounce, which we called as the promptly MHD exploding models.
 We furthermore find that even for the model with 
the weaker initial field and slow rotation, the jet-like explosions 
can occur after sufficient field wrapping to reach the critical field
strength, which we called as the delayed MHD exploding models. In this
case, the explosion can be delayed about $\sim 100\mathrm{ms}$ after
bounce. The explosion energy also strongly depends on
      the time difference before the shock-revival.
      The stronger initial magnetic fields make wider
exploding regions, leading to the larger explosion energy. The largest
MHD-driven explosion energy obtained is $\sim 10^{50}$ erg.
\end{itemize}


\textcolor{red}{In addition to the magnetic shock-revival,
 the neutrino-driven shock revival, namely neutrino heating
 from the newly born protoneutron star may energetize the jets 
as suggested by \citet{thom07}. 
Although we treated only the neutrino coolings in our computations, 
 we try to estimate the effect of the neutrino heating in the following way 
to see which one could be more important to produce the jets.
We compare the energy gained by the neutrino heating 
to the magnetic energy.
The specific neutrino heating rate due to neutrino absorptions 
($ \nu_e + n \rightarrow e^{-}+ p$ and ${\bar\nu}_e + p \rightarrow e^{+}+ n$),
 which are the dominant heating processes at certain radius in the postbounce phase,
can be estimated using Eq. (10) of \citet{qian96}.
For model B11TW1.0 at $22\ms$ after bounce,
the jet reaches $\sim 5\times 10^{7} \cm$ and the density at the head of the jet, 
$\rho_{\rm jet}$, is $\sim 1.5 \times 10^8 \gpcmc$.
At that time, the neutrino sphere locates at $\sim 7 \times 10^{6} \cm$.
The average energies and the luminosity for the electron and anti-electron neutrinos
 there are about $10$ and $14 \MeV$, and $50$ and $\sim 7 \times 10^{51} \ergps$.
In this setup, the heating rate due to the neutrino absorptions, $\dot{q}_{\nu N}$, 
reads $\sim 7.7 \times 10^{22} {\rm MeV}/{\rm s}/{\rm g}$. 
 In the same way, the heating rate due to the neutrino pair-annihilation 
($\nu + {\bar \nu} \rightarrow e^{-}+ e^{+}$) can be estimated as, 
$\sim 1.0 \times 10^{16} {\rm MeV}/{\rm s}/{\rm g}$, which is negligible
compared to the neutrino 
absorptions, albeit with the general relativistic corrections 
\citep{salm99,asano1,asano2}, for far outside the neutrino 
spheres where the magnetic shock revival occurs. 
If the neutrino luminosity maintains during the characteristic time 
scale for the delayed MHD explosion, namely $\Delta t_{\rm delayed} \sim 100$ ms, 
the fluid element may gain $\dot{q}_{\nu N}(\rho_{\rm jet}
/1.5 \times 10^8 \gpcmc)(\Delta t_{\rm delayed}/100{\rm ms}) \sim 1.6\times 10^{25}
\ergpcmc$ from the neutrino heatings.
This estimation shows that the energy deposition from neutrinos is quite smaller 
than the magnetic energy behind the jet,
$\sim \frac{(B/10^{15}{\rm G})^2}{8\pi}\sim 5.4\times 10^{28}\ergpcmc$.
Here, we omitted the neutrino cooling to maximize the effect of the neutrino heating.
 This simple estimation shows that for the models computed in this paper, the
 neutrino heatings could be minor compared to the magnetic effects. However for
 more weakly magnetized and slowly rotating cores, the neutrino heating may 
overwhelm the magnetic shock-revival. To seek the criteria which mechanisms can 
be dominant, is important, however, beyond scope of this study.}

{\color{red}
With respect to our numerical computations, we have to give some discussions.
First of all, we discuss the convergence of our numerical results. 
As mentioned in section \ref{sec:NM}, we have taken the grid numbers 
of $N_r = 300$ and $N_{\theta} = 60$ 
as a fiducial value with $N_r$ and $N_{\theta}$ being the grid numbers in radial and 
polar direction, respectively. For the convergence tests,  we vary $N_{r}$ from $200$, 
$400$, to $1000$ fixing $N_{\theta} =60$ and  $N_{\theta}$ from $30$, $90$, to $120$, 
while fixing  $N_{r} =300$. Taking model B11TW1.0, 
we follow the dynamics for models with the different numerical resolutions
till $\sim 30$ ms after bounce when the magneto-driven jets 
 come out of the central iron cores. 
We pay attention to the (spatially-integrated) magnetic energies 
among the models, because they are a good indicator to see the degree of 
the amplification of the magnetic fields.
Varying the coarser to the finer grid resolutions for the radial and polar direction, 
the relative change in the magnetic energies from the fiducial model is found to 
be in the range of $+0.15\%-0.35\%$ and $\pm 3.0\%$, respectively.
These resolution tests seem to support the convergence of the numerical values
 obtained in this paper. However we shall note that 
 better angular resolutions are needed to resolve the collimation of the jets 
further than $\sim 4\times 10^7$ cm from the center. 
This is because the intervals of the radial mesh, $\Delta r$, become logarithmically large in the spherical coordinates, 
which may lead to the overestimation of the collimation.
Since the jets propagate rather cylindrically around the polar axis later on, 
we think it a good way to switch from 
the spherical coordinates to the cylindrical coordinates at certain radius 
to sustain better angular resolutions, however, beyond scope of this study.
Next we estimate the numerical dissipation of the magnetic fields.
During the infall epoch, the toroidal magnetic flux is a good quantity to 
see the flux conservation. The difference between the initial toroidal magnetic flux, 
$\Phi_{\mathrm{ini}}$, and the flux at the bounce, $\Phi_{\mathrm{bnc}}$, is
estimated to be $\Delta\Phi = \frac{\Phi_{\mathrm{ini}}-\Phi_{\mathrm{bnc}}}{\Phi_{\mathrm{ini}}} \sim 1 \%$.
 This means that the numerical magnetic dissipation is 
 treated to be small for our numerical code.}

Then we move on to address a few imperfections in our simulations.
Magneto-rotational Instability (MRI) has been pointed out to be active 
in the outer layer of protoneutron star
which rotates strongly differentially \citep{akiy03,kota04b,masa06}.
The wavelength of maximum growth rate of the linear instability,
$\lambda=\frac{2\pi v_{a}}{\Omega}$, becomes 
$
\sim 5
\left(\frac{300 \mathrm{s}^{-1}}{\Omega}\right)
\left(\frac{B}{10^{12}\mathrm{G}}\right)
\left(\frac{10^{10}\mathrm{g/cm^3}}{\rho}\right)^{1/2}\mathrm{km}
$
\citep{balb98}, where our numerical grid is $\sim 3\mathrm{km}$
 there. Since $10-100$ times finer mesh than the wavelength is
required for resolving the MRI \citep{shib06}, our simulations are
insufficient to take into account the field amplification due to the MRI. 
This is a very important task remained to be investigated, 
although the computational costs are still expensive.  
As for the microphysics, the neutrino heating is not included as mentioned above 
(see, however, \citet{burr07}). Since the timescale for the magnetic explosions in
the polar direction are much earlier than the neutrino heating or 
the g-mode excited explosions, we think that both of them basically 
should play a minor role here. 
On the other hand, the neutrino
heating could be helpful for producing the explosion in the direction
of the equatorial plane where the field-wrapping induced 
magnetic explosions are unlikely to occur. 
Finally the general relativistic (GR) effects are
not incorporated in our simulations. 
During our simulation time, 
the central protoneutron stars do not collapse to the black
holes as inferred from a simple estimation of the Schwarzshild radius.
Thus we think that the GR effects may not drastically change our
results qualitatively.  
\textcolor{red}{
After the jet break-out from the star leaving behind the 
 narrow funnel, the mass accretion in the direction of the equatorial plane 
 may lead to the black hole formation, which provides us with 
 the initial condition of the collapsar. This would be another interesting 
subject as an application of this study.}

Bearing these caveats in mind,
we state some speculations based on our results.
The protoneutron stars obtained here are with the poloidal magnetic fields of order $10^{15}[\rm{G}]$ and with its
 rotation period of an order of milliseconds, which could be the
origin of the magnetar according to a hypothesis by 
 \citet{dunc92}.
If so, it means that the magnetically-driven jets could be
associated with the formation of the magnetars.
Our results suggest that the toroidal component of the magnetic field
is dominant in the young magnetars.
The large magnetic energy of the toroidal component stored in deep crusts
and cores of the magnetars would be transported outside via \Alfven waves
and be released as giant-flares of the SGRs \citep{dunc92,thom01,thom02}.
Some observational evidences support this picture.
For example, quasi periodic oscillations discovered in a X-ray light
curve of the giant-flare from SGR1806-20 would originate from the \Alfven
wave from the interior of the star \citep{rea06,isra05}. 
There are several studies indicating that the magnetar formation yields
the XRF \citep{maza06,maed07a,maed07b}.
While the ordinary GRBs require the highly relativistic ejecta,
the mildly relativistic ejecta is favorable for XRFs
\citep{toma07,sode06,ghis07}, which may be the case here because 
the magneto-driven jets become only mildly relativistic due 
to the high baryon loading of the matter along the rotational axis.

There remain more rooms to be investigated applying our simulations.
In this study, we employed one progenitor model. Since
the accretion rate of the matter to the protoneutron stars should 
depend on the progenitor mass \citep{hege05,woos06}, we think it very
important to investigate how the criterion of
the magnetic shock-revival changes with the progenitor models.
Moreover the initial configurations of the magnetic fields, which are
still highly uncertain, could be
changed in a systematic manner like in \citet{sawa05,sawa07}, to see
their effects on dynamics.
\textcolor{red}{While this study focused on the shock-propagation in the iron cores, 
 in which the jets become only mildly relativistic,
 we plan to continue to follow the dynamics later on till the jets break out of the stars
 (phase 4 in the introduction), in which the jets are expected to be relativistic.}
Very recent studies by \citet{komi07b,bucc07} are on this line. 
Our simulation can be more consistent than their studies in the sense 
that we start the simulations from the onset of the core-collapse, and 
that the protoneutron stars are not excised like their models.
By continuing the simulations of the jet propagations till the
shock-breakout, we plan to study the possible connection between the
magnetically-driven jets obtained here and the origins of the XRFs 
in the forthcoming work (Takiwaki et al. in preparation). 
\acknowledgments

We are grateful to S. Yamada for fruitful discussions.
KK thanks to E. M\"{uller} for useful discussions.
TT is grateful to S. Horiuchi for proofreading the manuscript.
Numerical computations were in part carried on VPP5000
 and general common use computer system at the center for
Computational Astrophysics, CfCA, the National
Astronomical Observatory of Japan.
This study was partially supported
by Grants-in-Aid for the Scientific Research from the Ministry of Education, Science and
Culture of Japan through No.S 19104006, No. 14079202 and No. 16740134.
This work was supported by World Premier International  Research Center Initiative (WPI Initiative), MEXT, Japan.
\bibliographystyle{apj}
\bibliography{sr}

\begin{thebibliography}{105}
\expandafter\ifx\csname natexlab\endcsname\relax\def\natexlab#1{#1}\fi

\bibitem[{{Akiyama} {et~al.}(2003){Akiyama}, {Wheeler}, {Meier}, \&
  {Lichtenstadt}}]{akiy03}
{Akiyama}, S., {Wheeler}, J.~C., {Meier}, D.~L., \& {Lichtenstadt}, I. 2003,
  \apj, 584, 954

\bibitem[{{Aloy} {et~al.}(2000){Aloy}, {M{\"u}ller}, {Ib{\'a}{\~n}ez},
  {Mart{\'{\i}}}, \& {MacFadyen}}]{aloy00}
{Aloy}, M.~A., {M{\"u}ller}, E., {Ib{\'a}{\~n}ez}, J.~M., {Mart{\'{\i}}},
  J.~M., \& {MacFadyen}, A. 2000, \apjl, 531, L119

\bibitem[{{Anninos} {et~al.}(2003){Anninos}, {Fragile}, \& {Murray}}]{anni03}
{Anninos}, P., {Fragile}, P.~C., \& {Murray}, S.~D. 2003, \apjs, 147, 177

\bibitem[{{Asano} \& {Fukuyama}(2000)}]{asano2}
{Asano}, K. \& {Fukuyama}, T. 2000, \apj, 531, 949

\bibitem[{{Asano} \& {Fukuyama}(2001)}]{asano1}
---. 2001, \apj, 546, 1019

\bibitem[{{Balbus} \& {Hawley}(1998)}]{balb98}
{Balbus}, S.~A. \& {Hawley}, J.~F. 1998, Reviews of Modern Physics, 70, 1

\bibitem[{{Barkov} \& {Komissarov}(2008)}]{bark08}
{Barkov}, M.~V. \& {Komissarov}, S.~S. 2008, \mnras, 385, L28

\bibitem[{{Brio} \& {Wu}(1988)}]{brio88}
{Brio}, M. \& {Wu}, C.~C. 1988, Journal of Computational Physics, 75, 400

\bibitem[{{Bucciantini} {et~al.}(2007){Bucciantini}, {Quataert}, {Arons},
  {Metzger}, \& {Thompson}}]{bucc07}
{Bucciantini}, N., {Quataert}, E., {Arons}, J., {Metzger}, B.~D., \&
  {Thompson}, T.~A. 2007, \mnras, L126+

\bibitem[{{Burrows} {et~al.}(2007){Burrows}, {Dessart}, {Livne}, {Ott}, \&
  {Murphy}}]{burr07}
{Burrows}, A., {Dessart}, L., {Livne}, E., {Ott}, C.~D., \& {Murphy}, J. 2007,
  \apj, 664, 416

\bibitem[{{Cerd{\'a}-Dur{\'a}n} {et~al.}(2007){Cerd{\'a}-Dur{\'a}n}, {Font}, \&
  {Dimmelmeier}}]{cerd07}
{Cerd{\'a}-Dur{\'a}n}, P., {Font}, J.~A., \& {Dimmelmeier}, H. 2007, ArXiv
  Astrophysics e-prints

\bibitem[{{De Villiers} {et~al.}(2003){De Villiers}, {Hawley}, \&
  {Krolik}}]{devi03}
{De Villiers}, J.-P., {Hawley}, J.~F., \& {Krolik}, J.~H. 2003, \apj, 599, 1238

\bibitem[{{De Villiers} {et~al.}(2005){De Villiers}, {Hawley}, {Krolik}, \&
  {Hirose}}]{devi05}
{De Villiers}, J.-P., {Hawley}, J.~F., {Krolik}, J.~H., \& {Hirose}, S. 2005,
  \apj, 620, 878

\bibitem[{{Duncan} \& {Thompson}(1992)}]{dunc92}
{Duncan}, R.~C. \& {Thompson}, C. 1992, \apjl, 392, L9

\bibitem[{{Epstein} \& {Pethick}(1981)}]{epst81}
{Epstein}, R.~I. \& {Pethick}, C.~J. 1981, \apj, 243, 1003

\bibitem[{{Fujimoto} {et~al.}(2006){Fujimoto}, {Kotake}, {Yamada}, {Hashimoto},
  \& {Sato}}]{fuji06}
{Fujimoto}, S.-i., {Kotake}, K., {Yamada}, S., {Hashimoto}, M.-a., \& {Sato},
  K. 2006, \apj, 644, 1040

\bibitem[{{Fuller} {et~al.}(1985){Fuller}, {Fowler}, \& {Newman}}]{full85}
{Fuller}, G.~M., {Fowler}, W.~A., \& {Newman}, M.~J. 1985, \apj, 293, 1

\bibitem[{{Galama} {et~al.}(1998){Galama}, {Vreeswijk}, {van Paradijs},
  {Kouveliotou}, {Augusteijn}, {B{\"o}hnhardt}, {Brewer}, {Doublier},
  {Gonzalez}, {Leibundgut}, {Lidman}, {Hainaut}, {Patat}, {Heise}, {in't Zand},
  {Hurley}, {Groot}, {Strom}, {Mazzali}, {Iwamoto}, {Nomoto}, {Umeda},
  {Nakamura}, {Young}, {Suzuki}, {Shigeyama}, {Koshut}, {Kippen}, {Robinson},
  {de Wildt}, {Wijers}, {Tanvir}, {Greiner}, {Pian}, {Palazzi}, {Frontera},
  {Masetti}, {Nicastro}, {Feroci}, {Costa}, {Piro}, {Peterson}, {Tinney},
  {Boyle}, {Cannon}, {Stathakis}, {Sadler}, {Begam}, \& {Ianna}}]{gala98}
{Galama}, T.~J., {Vreeswijk}, P.~M., {van Paradijs}, J., {Kouveliotou}, C.,
  {Augusteijn}, T., {B{\"o}hnhardt}, H., {Brewer}, J.~P., {Doublier}, V.,
  {Gonzalez}, J.-F., {Leibundgut}, B., {Lidman}, C., {Hainaut}, O.~R., {Patat},
  F., {Heise}, J., {in't Zand}, J., {Hurley}, K., {Groot}, P.~J., {Strom},
  R.~G., {Mazzali}, P.~A., {Iwamoto}, K., {Nomoto}, K., {Umeda}, H.,
  {Nakamura}, T., {Young}, T.~R., {Suzuki}, T., {Shigeyama}, T., {Koshut}, T.,
  {Kippen}, M., {Robinson}, C., {de Wildt}, P., {Wijers}, R.~A.~M.~J.,
  {Tanvir}, N., {Greiner}, J., {Pian}, E., {Palazzi}, E., {Frontera}, F.,
  {Masetti}, N., {Nicastro}, L., {Feroci}, M., {Costa}, E., {Piro}, L.,
  {Peterson}, B.~A., {Tinney}, C., {Boyle}, B., {Cannon}, R., {Stathakis}, R.,
  {Sadler}, E., {Begam}, M.~C., \& {Ianna}, P. 1998, \nat, 395, 670

\bibitem[{{Gavriil} {et~al.}(2002){Gavriil}, {Kaspi}, \& {Woods}}]{gavr02}
{Gavriil}, F.~P., {Kaspi}, V.~M., \& {Woods}, P.~M. 2002, \nat, 419, 142

\bibitem[{{Ghisellini} {et~al.}(2007){Ghisellini}, {Ghirlanda}, \&
  {Tavecchio}}]{ghis07}
{Ghisellini}, G., {Ghirlanda}, G., \& {Tavecchio}, F. 2007, \mnras, L117+

\bibitem[{{Harding} \& {Lai}(2006)}]{hard06}
{Harding}, A.~K. \& {Lai}, D. 2006, Reports of Progress in Physics, 69, 2631

\bibitem[{{Hawley} \& {Krolik}(2006)}]{hawl06}
{Hawley}, J.~F. \& {Krolik}, J.~H. 2006, \apj, 641, 103

\bibitem[{{Hawley} {et~al.}(1984{\natexlab{a}}){Hawley}, {Smarr}, \&
  {Wilson}}]{hawl84a}
{Hawley}, J.~F., {Smarr}, L.~L., \& {Wilson}, J.~R. 1984{\natexlab{a}}, \apj,
  277, 296

\bibitem[{{Hawley} {et~al.}(1984{\natexlab{b}}){Hawley}, {Smarr}, \&
  {Wilson}}]{hawl84b}
---. 1984{\natexlab{b}}, \apjs, 55, 211

\bibitem[{{Heger} \& {Langer}(2000)}]{hege00}
{Heger}, A. \& {Langer}, N. 2000, \apj, 544, 1016

\bibitem[{{Heger} {et~al.}(2005){Heger}, {Woosley}, \& {Spruit}}]{hege05}
{Heger}, A., {Woosley}, S.~E., \& {Spruit}, H.~C. 2005, \apj, 626, 350

\bibitem[{{Ibrahim} {et~al.}(2003){Ibrahim}, {Swank}, \& {Parke}}]{ibra03}
{Ibrahim}, A.~I., {Swank}, J.~H., \& {Parke}, W. 2003, \apjl, 584, L17

\bibitem[{{Israel} {et~al.}(2005){Israel}, {Belloni}, {Stella}, {Rephaeli},
  {Gruber}, {Casella}, {Dall'Osso}, {Rea}, {Persic}, \& {Rothschild}}]{isra05}
{Israel}, G.~L., {Belloni}, T., {Stella}, L., {Rephaeli}, Y., {Gruber}, D.~E.,
  {Casella}, P., {Dall'Osso}, S., {Rea}, N., {Persic}, M., \& {Rothschild},
  R.~E. 2005, \apjl, 628, L53

\bibitem[{{Itoh} {et~al.}(1989){Itoh}, {Adachi}, {Nakagawa}, {Kohyama}, \&
  {Munakata}}]{itoh89}
{Itoh}, N., {Adachi}, T., {Nakagawa}, M., {Kohyama}, Y., \& {Munakata}, H.
  1989, \apj, 339, 354

\bibitem[{{Itoh} {et~al.}(1990){Itoh}, {Adachi}, {Nakagawa}, {Kohyama}, \&
  {Munakata}}]{itoh90}
---. 1990, \apj, 360, 741

\bibitem[{{Janka} {et~al.}(2007){Janka}, {Langanke}, {Marek},
  {Mart{\'{\i}}nez-Pinedo}, \& {M{\"u}ller}}]{jank07}
{Janka}, H.-T., {Langanke}, K., {Marek}, A., {Mart{\'{\i}}nez-Pinedo}, G., \&
  {M{\"u}ller}, B. 2007, \physrep, 442, 38

\bibitem[{{Koide} {et~al.}(1998){Koide}, {Shibata}, \& {Kudoh}}]{koid98}
{Koide}, S., {Shibata}, K., \& {Kudoh}, T. 1998, \apjl, 495, L63+

\bibitem[{{Komissarov} \& {Barkov}(2007)}]{komi07b}
{Komissarov}, S.~S. \& {Barkov}, M.~V. 2007, \mnras, 382, 1029

\bibitem[{{Komissarov} \& {McKinney}(2007)}]{komi07a}
{Komissarov}, S.~S. \& {McKinney}, J.~C. 2007, \mnras, 377, L49

\bibitem[{{Kotake} {et~al.}(2006){Kotake}, {Sato}, \& {Takahashi}}]{kota06}
{Kotake}, K., {Sato}, K., \& {Takahashi}, K. 2006, Reports of Progress in
  Physics, 69, 971

\bibitem[{{Kotake} {et~al.}(2004{\natexlab{a}}){Kotake}, {Sawai}, {Yamada}, \&
  {Sato}}]{kota04b}
{Kotake}, K., {Sawai}, H., {Yamada}, S., \& {Sato}, K. 2004{\natexlab{a}},
  \apj, 608, 391

\bibitem[{{Kotake} {et~al.}(2003){Kotake}, {Yamada}, \& {Sato}}]{kota03a}
{Kotake}, K., {Yamada}, S., \& {Sato}, K. 2003, \apj, 595, 304

\bibitem[{{Kotake} {et~al.}(2004{\natexlab{b}}){Kotake}, {Yamada}, {Sato},
  {Sumiyoshi}, {Ono}, \& {Suzuki}}]{kota04a}
{Kotake}, K., {Yamada}, S., {Sato}, K., {Sumiyoshi}, K., {Ono}, H., \&
  {Suzuki}, H. 2004{\natexlab{b}}, \prd, 69, 124004

\bibitem[{{Krolik} {et~al.}(2005){Krolik}, {Hawley}, \& {Hirose}}]{krol05}
{Krolik}, J.~H., {Hawley}, J.~F., \& {Hirose}, S. 2005, \apj, 622, 1008

\bibitem[{{Lattimer} \& {Prakash}(2007)}]{latt07}
{Lattimer}, J.~M. \& {Prakash}, M. 2007, \physrep, 442, 109

\bibitem[{{LeBlanc} \& {Wilson}(1970)}]{leblanc}
{LeBlanc}, J.~M. \& {Wilson}, J.~R. 1970, \apj, 161, 541

\bibitem[{{Leismann} {et~al.}(2005){Leismann}, {Ant{\'o}n}, {Aloy},
  {M{\"u}ller}, {Mart{\'{\i}}}, {Miralles}, \& {Ib{\'a}{\~n}ez}}]{leis05}
{Leismann}, T., {Ant{\'o}n}, L., {Aloy}, M.~A., {M{\"u}ller}, E.,
  {Mart{\'{\i}}}, J.~M., {Miralles}, J.~A., \& {Ib{\'a}{\~n}ez}, J.~M. 2005,
  \aap, 436, 503

\bibitem[{{MacFadyen} \& {Woosley}(1999)}]{macf99}
{MacFadyen}, A.~I. \& {Woosley}, S.~E. 1999, \apj, 524, 262

\bibitem[{{Maeda} {et~al.}(2007{\natexlab{a}}){Maeda}, {Kawabata}, {Tanaka},
  {Nomoto}, {Tominaga}, {Hattori}, {Minezaki}, {Kuroda}, {Suzuki}, {Deng},
  {Mazzali}, \& {Pian}}]{maed07a}
{Maeda}, K., {Kawabata}, K., {Tanaka}, M., {Nomoto}, K., {Tominaga}, N.,
  {Hattori}, T., {Minezaki}, T., {Kuroda}, T., {Suzuki}, T., {Deng}, J.,
  {Mazzali}, P.~A., \& {Pian}, E. 2007{\natexlab{a}}, \apjl, 658, L5

\bibitem[{{Maeda} {et~al.}(2007{\natexlab{b}}){Maeda}, {Tanaka}, {Nomoto},
  {Tominaga}, {Kawabata}, {Mazzali}, {Umeda}, {Suzuki}, \& {Hattori}}]{maed07b}
{Maeda}, K., {Tanaka}, M., {Nomoto}, K., {Tominaga}, N., {Kawabata}, K.,
  {Mazzali}, P.~A., {Umeda}, H., {Suzuki}, T., \& {Hattori}, T.
  2007{\natexlab{b}}, \apj, 666, 1069

\bibitem[{{Masada} {et~al.}(2006){Masada}, {Sano}, \& {Takabe}}]{masa06}
{Masada}, Y., {Sano}, T., \& {Takabe}, H. 2006, \apj, 641, 447

\bibitem[{{Matt} {et~al.}(2006){Matt}, {Frank}, \& {Blackman}}]{matt06}
{Matt}, S., {Frank}, A., \& {Blackman}, E.~G. 2006, \apjl, 647, L45

\bibitem[{{Mazzali} {et~al.}(2006){Mazzali}, {Deng}, {Nomoto}, {Sauer}, {Pian},
  {Tominaga}, {Tanaka}, {Maeda}, \& {Filippenko}}]{maza06}
{Mazzali}, P.~A., {Deng}, J., {Nomoto}, K., {Sauer}, D.~N., {Pian}, E.,
  {Tominaga}, N., {Tanaka}, M., {Maeda}, K., \& {Filippenko}, A.~V. 2006, \nat,
  442, 1018

\bibitem[{{McKinney}(2006)}]{mcki06}
{McKinney}, J.~C. 2006, \mnras, 368, 1561

\bibitem[{{McKinney} \& {Narayan}(2007)}]{mcki07b}
{McKinney}, J.~C. \& {Narayan}, R. 2007, \mnras, 375, 531

\bibitem[{{Meier} {et~al.}(1976){Meier}, {Epstein}, {Arnett}, \&
  {Schramm}}]{meier}
{Meier}, D.~L., {Epstein}, R.~I., {Arnett}, W.~D., \& {Schramm}, D.~N. 1976,
  \apj, 204, 869

\bibitem[{{Meszaros}(2006)}]{mesz06}
{Meszaros}, P. 2006, Reports of Progress in Physics, 69, 2259

\bibitem[{{Metzger} {et~al.}(2007){Metzger}, {Thompson}, \&
  {Quataert}}]{thom07}
{Metzger}, B.~D., {Thompson}, T.~A., \& {Quataert}, E. 2007, \apj, 659, 561

\bibitem[{{Mizuno} {et~al.}(2007){Mizuno}, {Hardee}, \& {Nishikawa}}]{mizu07}
{Mizuno}, Y., {Hardee}, P., \& {Nishikawa}, K.-I. 2007, \apj, 662, 835

\bibitem[{{Mizuno} {et~al.}(2006){Mizuno}, {Nishikawa}, {Koide}, {Hardee}, \&
  {Fishman}}]{mizu06}
{Mizuno}, Y., {Nishikawa}, K.-I., {Koide}, S., {Hardee}, P., \& {Fishman},
  G.~J. 2006, ArXiv Astrophysics e-prints

\bibitem[{{Mizuta} {et~al.}(2006){Mizuta}, {Yamasaki}, {Nagataki}, \&
  {Mineshige}}]{mizt06}
{Mizuta}, A., {Yamasaki}, T., {Nagataki}, S., \& {Mineshige}, S. 2006, \apj,
  651, 960

\bibitem[{{Moiseenko} {et~al.}(2006){Moiseenko}, {Bisnovatyi-Kogan}, \&
  {Ardeljan}}]{mois06}
{Moiseenko}, S.~G., {Bisnovatyi-Kogan}, G.~S., \& {Ardeljan}, N.~V. 2006,
  \mnras, 370, 501

\bibitem[{{Nagataki} {et~al.}(2007){Nagataki}, {Takahashi}, {Mizuta}, \&
  {Takiwaki}}]{naga07}
{Nagataki}, S., {Takahashi}, R., {Mizuta}, A., \& {Takiwaki}, T. 2007, \apj,
  659, 512

\bibitem[{{Nishikawa} {et~al.}(2005){Nishikawa}, {Richardson}, {Koide},
  {Shibata}, {Kudoh}, {Hardee}, \& {Fishman}}]{nisi05}
{Nishikawa}, K.-I., {Richardson}, G., {Koide}, S., {Shibata}, K., {Kudoh}, T.,
  {Hardee}, P., \& {Fishman}, G.~J. 2005, \apj, 625, 60

\bibitem[{{Nishimura} {et~al.}(2006){Nishimura}, {Kotake}, {Hashimoto},
  {Yamada}, {Nishimura}, {Fujimoto}, \& {Sato}}]{nish06}
{Nishimura}, S., {Kotake}, K., {Hashimoto}, M.-a., {Yamada}, S., {Nishimura},
  N., {Fujimoto}, S., \& {Sato}, K. 2006, \apj, 642, 410

\bibitem[{{Obergaulinger} {et~al.}(2006){Obergaulinger}, {Aloy}, {Dimmelmeier},
  \& {M{\"u}ller}}]{ober06a}
{Obergaulinger}, M., {Aloy}, M.~A., {Dimmelmeier}, H., \& {M{\"u}ller}, E.
  2006, \aap, 457, 209

\bibitem[{{Paczynski}(1998)}]{pacz98}
{Paczynski}, B. 1998, \apjl, 494, L45+

\bibitem[{{Piran}(1999)}]{pira99}
{Piran}, T. 1999, \physrep, 314, 575

\bibitem[{{Piro} {et~al.}(1998){Piro}, {Amati}, {Antonelli}, {Butler}, {Costa},
  {Cusumano}, {Feroci}, {Frontera}, {Heise}, {in 't Zand}, {Molendi}, {Muller},
  {Nicastro}, {Orlandini}, {Owens}, {Parmar}, {Soffitta}, \& {Tavani}}]{piro98}
{Piro}, L., {Amati}, L., {Antonelli}, L.~A., {Butler}, R.~C., {Costa}, E.,
  {Cusumano}, G., {Feroci}, M., {Frontera}, F., {Heise}, J., {in 't Zand},
  J.~J.~M., {Molendi}, S., {Muller}, J., {Nicastro}, L., {Orlandini}, M.,
  {Owens}, A., {Parmar}, A.~N., {Soffitta}, P., \& {Tavani}, M. 1998, \aap,
  331, L41

\bibitem[{{Proga} {et~al.}(2003){Proga}, {MacFadyen}, {Armitage}, \&
  {Begelman}}]{prog03}
{Proga}, D., {MacFadyen}, A.~I., {Armitage}, P.~J., \& {Begelman}, M.~C. 2003,
  \apjl, 599, L5

\bibitem[{{Qian} \& {Woosley}(1996)}]{qian96}
{Qian}, Y.-Z. \& {Woosley}, S.~E. 1996, \apj, 471, 331

\bibitem[{{Rea} {et~al.}(2006){Rea}, {Israel}, {Mereghetti}, {Tiengo}, {Zane},
  {Turolla}, \& {Stella}}]{rea06}
{Rea}, N., {Israel}, G.~L., {Mereghetti}, S., {Tiengo}, A., {Zane}, S.,
  {Turolla}, R., \& {Stella}, L. 2006, Chinese Journal of Astronomy and
  Astrophysics Supplement, 6, 155

\bibitem[{{Rea} {et~al.}(2003){Rea}, {Israel}, {Stella}, {Oosterbroek},
  {Mereghetti}, {Angelini}, {Campana}, \& {Covino}}]{rea03}
{Rea}, N., {Israel}, G.~L., {Stella}, L., {Oosterbroek}, T., {Mereghetti}, S.,
  {Angelini}, L., {Campana}, S., \& {Covino}, S. 2003, \apjl, 586, L65

\bibitem[{{Rosswog} \& {Liebend{\"o}rfer}(2003)}]{ross03}
{Rosswog}, S. \& {Liebend{\"o}rfer}, M. 2003, \mnras, 342, 673

\bibitem[{{Salmonson} \& {Wilson}(1999)}]{salm99}
{Salmonson}, J.~D. \& {Wilson}, J.~R. 1999, \apj, 517, 859

\bibitem[{{Savaglio} {et~al.}(2006){Savaglio}, {Glazebrook}, \& {Le
  Borgne}}]{sava06}
{Savaglio}, S., {Glazebrook}, K., \& {Le Borgne}, D. 2006, in American
  Institute of Physics Conference Series, Vol. 836, Gamma-Ray Bursts in the
  Swift Era, ed. S.~S. {Holt}, N.~{Gehrels}, \& J.~A. {Nousek}, 540--545

\bibitem[{{Sawai} {et~al.}(2005){Sawai}, {Kotake}, \& {Yamada}}]{sawa05}
{Sawai}, H., {Kotake}, K., \& {Yamada}, S. 2005, \apj, 631, 446

\bibitem[{{Sawai} {et~al.}(2007){Sawai}, {Kotake}, \& {Yamada}}]{sawa07}
---. 2007, ArXiv, 709.1795

\bibitem[{{Scheidegger} {et~al.}(2007){Scheidegger}, {Fischer}, \&
  {Liebendoerfer}}]{scei07}
{Scheidegger}, S., {Fischer}, T., \& {Liebendoerfer}, M. 2007, ArXiv e-prints,
  709.0168

\bibitem[{{Sekiguchi} \& {Shibata}(2007)}]{seki07}
{Sekiguchi}, Y. \& {Shibata}, M. 2007, Progress of Theoretical Physics, 117,
  1029

\bibitem[{{Shapiro} \& {Teukolsky}(1983)}]{shap83}
{Shapiro}, S.~L. \& {Teukolsky}, S.~A. 1983, {Black holes, white dwarfs, and
  neutron stars: The physics of compact objects} (Research supported by the
  National Science Foundation.~New York, Wiley-Interscience, 1983, 663 p.)

\bibitem[{{Shen} {et~al.}(1998){Shen}, {Toki}, {Oyamatsu}, \&
  {Sumiyoshi}}]{shen98}
{Shen}, H., {Toki}, H., {Oyamatsu}, K., \& {Sumiyoshi}, K. 1998, Nuclear
  Physics A, 637, 435

\bibitem[{{Shibata} {et~al.}(2006){Shibata}, {Liu}, {Shapiro}, \&
  {Stephens}}]{shib06}
{Shibata}, M., {Liu}, Y.~T., {Shapiro}, S.~L., \& {Stephens}, B.~C. 2006, \prd,
  74, 104026

\bibitem[{{Sod}(1978)}]{sod78}
{Sod}, G.~A. 1978, Journal of Computational Physics, 27, 1

\bibitem[{{Soderberg} {et~al.}(2006){Soderberg}, {Kulkarni}, {Nakar}, {Berger},
  {Cameron}, {Fox}, {Frail}, {Gal-Yam}, {Sari}, {Cenko}, {Kasliwal},
  {Chevalier}, {Piran}, {Price}, {Schmidt}, {Pooley}, {Moon}, {Penprase},
  {Ofek}, {Rau}, {Gehrels}, {Nousek}, {Burrows}, {Persson}, \&
  {McCarthy}}]{sode06}
{Soderberg}, A.~M., {Kulkarni}, S.~R., {Nakar}, E., {Berger}, E., {Cameron},
  P.~B., {Fox}, D.~B., {Frail}, D., {Gal-Yam}, A., {Sari}, R., {Cenko}, S.~B.,
  {Kasliwal}, M., {Chevalier}, R.~A., {Piran}, T., {Price}, P.~A., {Schmidt},
  B.~P., {Pooley}, G., {Moon}, D.-S., {Penprase}, B.~E., {Ofek}, E., {Rau}, A.,
  {Gehrels}, N., {Nousek}, J.~A., {Burrows}, D.~N., {Persson}, S.~E., \&
  {McCarthy}, P.~J. 2006, \nat, 442, 1014

\bibitem[{{Spruit}(2002)}]{spru02}
{Spruit}, H.~C. 2002, \aap, 381, 923

\bibitem[{{Stanek} {et~al.}(2006){Stanek}, {Gnedin}, {Beacom}, {Gould},
  {Johnson}, {Kollmeier}, {Modjaz}, {Pinsonneault}, {Pogge}, \&
  {Weinberg}}]{stan06}
{Stanek}, K.~Z., {Gnedin}, O.~Y., {Beacom}, J.~F., {Gould}, A.~P., {Johnson},
  J.~A., {Kollmeier}, J.~A., {Modjaz}, M., {Pinsonneault}, M.~H., {Pogge}, R.,
  \& {Weinberg}, D.~H. 2006, Acta Astronomica, 56, 333

\bibitem[{{Stanek} {et~al.}(2003){Stanek}, {Matheson}, {Garnavich}, {Martini},
  {Berlind}, {Caldwell}, {Challis}, {Brown}, {Schild}, {Krisciunas}, {Calkins},
  {Lee}, {Hathi}, {Jansen}, {Windhorst}, {Echevarria}, {Eisenstein}, {Pindor},
  {Olszewski}, {Harding}, {Holland}, \& {Bersier}}]{stan03}
{Stanek}, K.~Z., {Matheson}, T., {Garnavich}, P.~M., {Martini}, P., {Berlind},
  P., {Caldwell}, N., {Challis}, P., {Brown}, W.~R., {Schild}, R.,
  {Krisciunas}, K., {Calkins}, M.~L., {Lee}, J.~C., {Hathi}, N., {Jansen},
  R.~A., {Windhorst}, R., {Echevarria}, L., {Eisenstein}, D.~J., {Pindor}, B.,
  {Olszewski}, E.~W., {Harding}, P., {Holland}, S.~T., \& {Bersier}, D. 2003,
  \apjl, 591, L17

\bibitem[{{Stone} \& {Hardee}(2000)}]{ston00}
{Stone}, J.~M. \& {Hardee}, P.~E. 2000, \apj, 540, 192

\bibitem[{{Stone} \& {Norman}(1992)}]{ston92}
{Stone}, J.~M. \& {Norman}, M.~L. 1992, \apjs, 80, 753

\bibitem[{{Suwa} {et~al.}(2007{\natexlab{a}}){Suwa}, {Takiwaki}, {Kotake}, \&
  {Sato}}]{suwa07b}
{Suwa}, Y., {Takiwaki}, T., {Kotake}, K., \& {Sato}, K. 2007{\natexlab{a}},
  \apjl, 665, L43

\bibitem[{{Suwa} {et~al.}(2007{\natexlab{b}}){Suwa}, {Takiwaki}, {Kotake}, \&
  {Sato}}]{suwa07a}
---. 2007{\natexlab{b}}, \pasj, 59, 771

\bibitem[{{Symbalisty}(1984)}]{symb84}
{Symbalisty}, E.~M.~D. 1984, \apj, 285, 729

\bibitem[{{Takahashi} {et~al.}(1978){Takahashi}, {El Eid}, \&
  {Hillebrandt}}]{taka78}
{Takahashi}, K., {El Eid}, M.~F., \& {Hillebrandt}, W. 1978, \aap, 67, 185

\bibitem[{{Takiwaki} {et~al.}(2004){Takiwaki}, {Kotake}, {Nagataki}, \&
  {Sato}}]{taki04}
{Takiwaki}, T., {Kotake}, K., {Nagataki}, S., \& {Sato}, K. 2004, \apj, 616,
  1086

\bibitem[{{Thompson} \& {Duncan}(2001)}]{thom01}
{Thompson}, C. \& {Duncan}, R.~C. 2001, \apj, 561, 980

\bibitem[{{Thompson} {et~al.}(2002){Thompson}, {Lyutikov}, \&
  {Kulkarni}}]{thom02}
{Thompson}, C., {Lyutikov}, M., \& {Kulkarni}, S.~R. 2002, \apj, 574, 332

\bibitem[{{Thompson} {et~al.}(2004){Thompson}, {Chang}, \& {Quataert}}]{thom04}
{Thompson}, T.~A., {Chang}, P., \& {Quataert}, E. 2004, \apj, 611, 380

\bibitem[{{Toma} {et~al.}(2007){Toma}, {Ioka}, {Sakamoto}, \&
  {Nakamura}}]{toma07}
{Toma}, K., {Ioka}, K., {Sakamoto}, T., \& {Nakamura}, T. 2007, \apj, 659, 1420

\bibitem[{{Usov}(1992)}]{usov92}
{Usov}, V.~V. 1992, \nat, 357, 472

\bibitem[{{Uzdensky} \& {MacFadyen}(2006)}]{uzde06}
{Uzdensky}, D.~A. \& {MacFadyen}, A.~I. 2006, \apj, 647, 1192

\bibitem[{{Uzdensky} \& {MacFadyen}(2007{\natexlab{a}})}]{uzde07a}
---. 2007{\natexlab{a}}, \apj, 669, 546

\bibitem[{{Uzdensky} \& {MacFadyen}(2007{\natexlab{b}})}]{uzde07b}
---. 2007{\natexlab{b}}, Physics of Plasmas, 14, 6506

\bibitem[{{van Leer}(1977)}]{vlee77}
{van Leer}, B. 1977, Journal of Computational Physics, 23, 276

\bibitem[{{Wheeler} {et~al.}(2000){Wheeler}, {Yi}, {H{\"o}flich}, \&
  {Wang}}]{whee00}
{Wheeler}, J.~C., {Yi}, I., {H{\"o}flich}, P., \& {Wang}, L. 2000, \apj, 537,
  810

\bibitem[{{Woosley} \& {Heger}(2006)}]{woos06}
{Woosley}, S.~E. \& {Heger}, A. 2006, \apj, 637, 914

\bibitem[{{Yamada} \& {Sawai}(2004)}]{yama04}
{Yamada}, S. \& {Sawai}, H. 2004, \apj, 608, 907

\bibitem[{{Yoon} \& {Langer}(2005)}]{yoon}
{Yoon}, S.-C. \& {Langer}, N. 2005, \aap, 443, 643

\bibitem[{{Zhang} \& {Harding}(2000)}]{zhan00}
{Zhang}, B. \& {Harding}, A.~K. 2000, \apjl, 535, L51

\bibitem[{{Zhang} {et~al.}(2003){Zhang}, {Woosley}, \& {MacFadyen}}]{zhan03}
{Zhang}, W., {Woosley}, S.~E., \& {MacFadyen}, A.~I. 2003, \apj, 586, 356

\end{thebibliography}

\appendix

\section{Derivation of the Basic Equations for SRMHD}\label{app:DeBE}
In this Appendix, we summarize formalisms on the basic equations and
the numerical tests for our newly developed SRMHD code.
For the formalisms, 
we follow the derivation of \citet{devi03,hawl84a,hawl84b}.
For convenience, we proceed
the derivation keeping the metric general forms, i.e.,
\textcolor{red}{ 
$\d s^2 = -\alpha^2 \d t^2 + \gamma_{ij}\left(dx^i + \beta^i dt \right)\left(dx^j + \beta^j dt \right)$} .
where $\alpha$
is the lapse function, $\beta$ is the
shift vector and $\gamma_{ij}$ is the spatial $3$-metric. 
And we take the Minkowski metric later.

There are four fundamental magnetohydrodynamic equations.
The conservation of baryon number is
\begin{eqnarray}
\udel{\mu}{\rho U^\mu}&=&0 \label{barcons}
\end{eqnarray}
where $\rho$,$U^{\mu}(\mu=0,1,2,3)$
are baryon mass density and 4-velocity at each point.
The conservation of the stress-energy is 
\begin{eqnarray}
\udel{\mu}{T^{\mu\nu}}&=&0 \label{tmncons}
\end{eqnarray}
where $T^{\mu\nu}$ is stress-energy tensor
and Maxwell's equations
\begin{eqnarray}
\udel{\mu}{F^{\mu\nu}}&=&4\pi J^{\nu},\\
\udel{\mu}{^{*}F^{\mu\nu}}&=&0\label{ind_eq}.
\end{eqnarray}
where $F^{\mu\nu}$ is the antisymmetric 
electro-magnetic tensor and 
\begin{eqnarray}
{^{*}F}^{\mu\nu}=\frac{1}{2}\epsilon^{\mu\nu\delta\sigma}F_{\delta\sigma}\label{fs_def}
\end{eqnarray}
is dual of $F^{\mu\nu}$.
Maxwell's equations are supplemented 
by the equation of the charge conservation $\udel{\mu}J^{\mu} = 0$.\\
The energy momentum tensor consists of perfect fluid parts
and electromagnetic parts, i.e.
\begin{eqnarray}
T^{\mu\nu}&=&\rho h^{*} U^{\mu}U^{\nu}+pg^{\mu\nu}
+\frac{1}{4\pi}\left({F^{\mu}}_{\alpha}F^{\nu\alpha}
-\frac{1}{4}F_{\alpha\beta}F^{\alpha\beta}g^{\mu\nu}
\right)
\end{eqnarray}
where $h^{*}=(1+e/\rho+p/\rho)$ is the relativistic enthalpy with $e$
and $p$ being the internal energy and the pressure, respectively.

For later convenience, 
we define  magnetic induction in 
the rest frame of the fluid,
\begin{eqnarray}
b^{\mu}&=&\frac{1}{\sqrt{4\pi}}{^{*}F^{\mu\nu}U_{\nu}}\label{b_def}.
\end{eqnarray}

We adopt the ideal MHD limit and assume infinite
conductivity (the flux- freezing condition),
where in the electric field in the fluid rest frame is zero, i.e., $F_{\mu\nu}U^{\nu}=0$.

Combining Eq. (\ref{fs_def}) with  (Eq. \ref{b_def})
and conditions for infinite conductivity, we obtain
\begin{equation}
 F_{\mu\nu} =\epsilon_{\alpha\beta\mu\nu}\sqrt{4\pi}b^{\alpha}U^{\beta}.
\end{equation}
The orthogonality condition 
\begin{equation}
 b^{\mu}U_{\mu}=0\label{b_on}
\end{equation}
follows directly from Eq. (\ref{b_def}).

The induction Eq. (\ref{ind_eq}) can also be 
rewritten by substituting the definitions,
\begin{equation}
\udel{\alpha}{}\left(U^\alpha\,b^\beta- b^\alpha\,U^\beta\right) =0 .
\end{equation}
By expanding this equation using the product rule and applying the 
orthogonality condition Eq. (\ref{b_on}), we obtain the identity
\begin{equation}
U_\nu\,b^\mu{\nabla}_{\mu}\,U^\nu =0 .\label{bident}
\end{equation}

It is useful to rewrite the energy momentum tensor as
\begin{eqnarray}
T^{\mu\nu}&=&
\left(\rho h^{*}+\left|b\right|^2 \right)U^{\mu}U^{\nu}
+(p+\frac{\left|b\right|^2}{2})g^{\mu\nu}
-b^{\mu}b^{\nu}.
\end{eqnarray}

We have to expand basic equations in terms of the code variable,
and transform the equation for auxiliary density, energy and momentum functions
$D=\rho W,E=eW,S_{i}=\rho h W^2 v_{i}$.
Finally the set of variables $D,E,S_{i},B_{i}$
will be evolved through the basic equations transformed here. 

The equation of baryon conservation (\ref{barcons}) 
can be expanded in terms of the code 
variables easily,
\begin{equation}\label{masscons}
\partial_t\,D + {1 \over
\sqrt{\gamma}}\,\partial_j\,(D\,\sqrt{\gamma}\,V^j)  = 0.
\end{equation}

The equation of energy conservation is derived by 
contracting (\ref{tmncons}) with ${U}_{\nu}$,
\begin{equation}\label{eq.1a}
{U}_{\nu}\,{\nabla}_{\mu}{T}^{\mu\,\nu} =
 {U}_{\nu}\,{\nabla}_{\mu}\left\{ 
 \left(\rho\,h^{*}+{\|b\|}^2\right)\,U^{\mu}\,U^{\nu}+
 \left(P+{{\|b\|}^2 \over 2}\right){g}^{\mu\,\nu}-
 b^\mu\,b^\nu\right\} = 0 .
\end{equation}
By using the identity (\ref{bident}) and 
(\ref{barcons}), we obtain the local energy conservation 
\begin{equation}\label{econs}
{\nabla}_{\mu} \left(\rho\,\epsilon\,U^{\mu}\right)+
 P\,{\nabla}_{\mu} U^{\mu}= 0 ,
\end{equation}
Applying the definition for the auxiliary energy function $E$, 
the energy equation is rewritten as follows:
\begin{equation} \label{enfinal}
 \partial_{t}\left(E\right)+{1 \over \sqrt{\gamma}}\,
\partial_{i}\left(\sqrt{\gamma}\,E\,V^i\right)
+ P\,\partial_{t}\left(W\right) + 
{P\over\sqrt{\gamma}}\,\partial_{i}\left(\sqrt{\gamma}\,W\,V^i\right)
= 0.
\end{equation}

The momentum conservation equations follow from
\begin{equation}
\nabla_\mu\,{T^\mu}_\nu = {\nabla}_{\mu}\left\{ 
 \left(\rho\,h^{*}+{\|b\|}^2\right)\,U^{\mu}\,U_{\nu}+
 \left(P+{{\|b\|}^2 \over 2}\right){\delta^\mu}_\nu-
 b^\mu\,b_\nu\right\} = 0 .\label{mom.1}
\end{equation}
This equation can be rewritten as
\begin{eqnarray}
& &{1 \over \alpha\,\sqrt{\gamma}}\,
\partial_\mu\,\sqrt{\gamma}\,S_\nu\,V^\mu +{1 \over 2\,\alpha}\,
{ S_\alpha\,S_\beta \over S^t }\,\partial_\nu\,g^{\alpha \beta}\nonumber\\
&+&
\partial_\nu\,\left(P+{{\|b\|}^2 \over 2}\right)-
{1 \over \alpha\,\sqrt{\gamma}}\,
\partial_\mu\,\alpha\,\sqrt{\gamma}\,b^\mu\,b_\nu-{1 \over 2}\,
 b_\alpha\,b_\beta\,\partial_\nu\,g^{\alpha \beta}
= 0 .\label{mom.2}
\end{eqnarray}
To obtain the final form of the equations, multiply (\ref{mom.2})
by the lapse $\alpha$, split the $\mu$ index into
its space ($i$) and time ($t$) components, and restrict $\nu$ to the
spatial indices ($j$) only:
\begin{eqnarray}
& &\partial_t\left(S_j-\alpha\,b_j\,b^t\right)+
  {1 \over \sqrt{\gamma}}\,
  \partial_i\,\sqrt{\gamma}\,\left(S_j\,V^i-\alpha\,b_j\,b^i\right)\nonumber\\
&=&
 - {1 \over 2}\,\left({S_\epsilon\,S_\mu \over S^t}+
  \alpha\,b_\mu\,b_\epsilon\right)\,
 \partial_j\,g^{\mu\,\epsilon}-
  \alpha\,\partial_j\left(P+{{\|b\|}^2 \over 2}\right)  .\label{mom.3}
\end{eqnarray}
The $\nu$ index can be restricted to the spatial indices because the
equation that arises from $\nu=t$ for the time components of momentum
and magnetic fields is redundant, corresponding to the total energy
conservation equation.
In our formalism, we solve the Eq.
(\ref{enfinal}) separately for the internal energy. 
Taking the following metric,
\begin{eqnarray}
g_{\mu\nu}&=&
\left(
\begin{array}{cc}
 -1& 0 \\
0  & \gamma_{ij} \\
\end{array}\right).
\end{eqnarray}
where $\gamma_{ij}$ is the spatial metric whose
concrete description depends on coordinate system.
Finally we describe our treatment of the gravity.
Under the weak field limit, time-time component of the metric, $g_{tt}$, takes
the form of $-\left(1-2\Phi\right)$ where
$\Phi$ is Newtonian gravitational potential \citep[e.g.][]{shap83}. 
The third term of the momentum equation then becomes,
\begin{eqnarray}
 -\frac{1}{2}\frac{S^{\alpha}S^{\beta}}{S^{t}}
\udel{j}g_{\alpha\beta}
\approx \rho h W^2 \udel{j}\Phi.
\end{eqnarray}
Under this limit,
Einstein equation becomes the Poisson equation for the 
gravitational potential (see Eq. (\ref{eq:poisson})).
Since the origin of the source term is
the $tt$ component of the energy momentum tensor,
we replace $\rho$ in the ordinary Newtonian limit with $T_{tt}$.
as we discussed in the appendix.

{{\color{red} The validity of using the Newtonian potential in the core-collapse 
simulations here may be discussed by seeing the 
value of compactness parameter, $\frac{2GM(r)}{rc^2}$, where $r$ is the radius and 
$M(r)$ is the enclosed mass
within $r$. In the vicinity of the protoneutron star of $\sim 1.2 M_{\odot}$ with 
the typical size of $\sim 20 {\rm km}$, the parameter is $\sim 0.18$. So 
the error caused by neglecting higher order metric perturbations 
is estimated to below $\sim 3\%$. The qualitative features 
found in this paper may be unchanged due to the incursion of the GR. }

\subsection{Energy Descriptions}\label{app:enedef}
We need to modify the description of energy from the Newtonian one to the special
relativistic one.
The total local energy, $e_{\mathrm{local}}$, is defined by sum of
the various energy:
\begin{eqnarray}
e_{\mathrm{local}}
&=&
 e_{\mathrm{kin}}
+e_{\mathrm{int}}
+e_{\mathrm{mag}}
+e_{\mathrm{grav}}
\end{eqnarray}
where
$e_{\mathrm{kin}},e_{\mathrm{int}},e_{\mathrm{grav}}$ and $e_{\mathrm{mag}}$ 
is kinetic energy, internal energy, gravitational energy 
and magnetic energy, respectively.
Their specific description is as follows:
\begin{eqnarray}
 e_{\mathrm{kin}}&=&\rho W \left(W-1\right),\\
 e_{\mathrm{int}}&=& e W^2+p\left(W^2-1\right),\\
 e_{\mathrm{grav}}&=&-\rho h W^2 \Phi,\\
 e_{\mathrm{mag}}&=& \vec{B}^2\left(1-\frac{1}{2W^2}\right)
-\frac{{b^0}^2}{2W^2}.
\end{eqnarray}
These descriptions are used for the calculations of the explosion
energy in Subsection \ref{sec:dependence}.

\section{Special Relativistic MOC}
The method of characteristics (MOC) is popularly used in the 
magneto-hydrodynamical simulations.
In this algorithm the magnetic fields are evolved along the
characteristic lines of 
the \Alfven waves.
Detailed procedure for this algorithm for the Newtonian case is given 
in \citet{ston92}. 
For the special relativistic (SR) computations, we derive the solutions of
 the SR \Alfven waves in an analytic form,
%
\begin{eqnarray}
\left.\Del{t}{Wv_i+b_i/\sqrt{\rho h}}\right|_{-}&=&0,\\
\left.\Del{t}{Wv_i-b_i/\sqrt{\rho h}}\right|_{+}&=&0,\\
\left.\Del{t}{}\right|_{-}
&\stackrel{\mathrm{def}}{\equiv}&\pdel{t}{}+\frac{v_j-\frac{b_j}{\sqrt{\rho h}W}}{(1-b^t/\sqrt{\rho h}W)}\pdel{x_j}{},\\
\left.\Del{t}{}\right|_{+}
&\stackrel{\mathrm{def}}{\equiv}&\pdel{t}{}+\frac{v_j+\frac{b_j}{\sqrt{\rho
h}W}}{(1+b^t/\sqrt{\rho h}W)}\pdel{x_j}{},
\end{eqnarray}
where $W$, $\rho$, and $h$ is the Lorentz factor, density and enthalpy respectively.
$v_j$ and $b_j$ is the perpendicular component of the velocity and the magnetic
field to the $x_{i}$ directions.  

In the subroutine for solving SR MOC in the code, the velocity and the magnetic fields
 are updated at half-time step along the characteristics using the above
equations. By giving the analytic forms, it is readily seen that the speed
 of the propagation is guaranteed to be below the speed of light even
 for the regions where the density becomes low and the magnetic fields become strong,
 which is quite important for keeping the stable numerical
 calculations in good accuracy.

\paragraph{\Alfven Wave Propagation}\label{ap:AWP}
The propagation of a liner \Alfven wave is a basic test problem 
of MHD simulation.
We consider a constant background magnetic field, $B_x$, and
fluid velocity, $v_{x}$.
And we add small transverse perturbations with velocity, 
$v_z$ ($v_y$), and magnetic field, $B_z$ ($B_y$).
In this situation
$b^t=\sum v^k b_k\approx v_j b_j$,
therefore the analytic solution for the \Alfven wave becomes
\begin{eqnarray}
\left.\Del{t}{Wv_z+b_z/\sqrt{\rho h}}\right|_{-}&=&0,\\
\left.\Del{t}{Wv_z-b_z/\sqrt{\rho h}}\right|_{+}&=&0,\\
\left.\Del{t}{}\right|_{-}
&\approx&\pdel{t}{}
+\frac{v_x-\frac{b_x}{\sqrt{Dh}W}}{1-v_x\frac{b_x}{\sqrt{Dh}W}}\pdel{x}{},\\
\left.\Del{t}{}\right|_{+}
&\approx&\pdel{t}{}
+\frac{v_x+\frac{b_x}{\sqrt{Dh}W}}{1+v_x\frac{b_x}{\sqrt{Dh}W}}\pdel{x}{}.
\end{eqnarray}
If we take 
\begin{eqnarray}
Wv_z+b_z/\sqrt{\rho h}=0, \label{nmmode}
\end{eqnarray}
the minus mode does not propagate.
we assume $B_x=0.09$, $v_x=0.08$
for the Newtonian \Alfven wave
and  $B_x=0.9$, $v_x=0.8$
for the relativistic \Alfven wave.
We take $v_z$ as $10^{-7}v_x\sin(2\pi x)$
and $B_z$ is determined from $b_z$ in Eq. (\ref{nmmode}).
The result is shown in Figure \ref{relalz}.
In both Newtonian and relativistic cases, the form of the wave is not changed.
It indicates that the computations are successfully performed in our code.
The propagated waveforms are very smooth and no oscillations are found
like the ones in the previous study \citep{devi03}.

\begin{figure}[ht]
 \begin{center}
   \includegraphics[width=.99\linewidth]{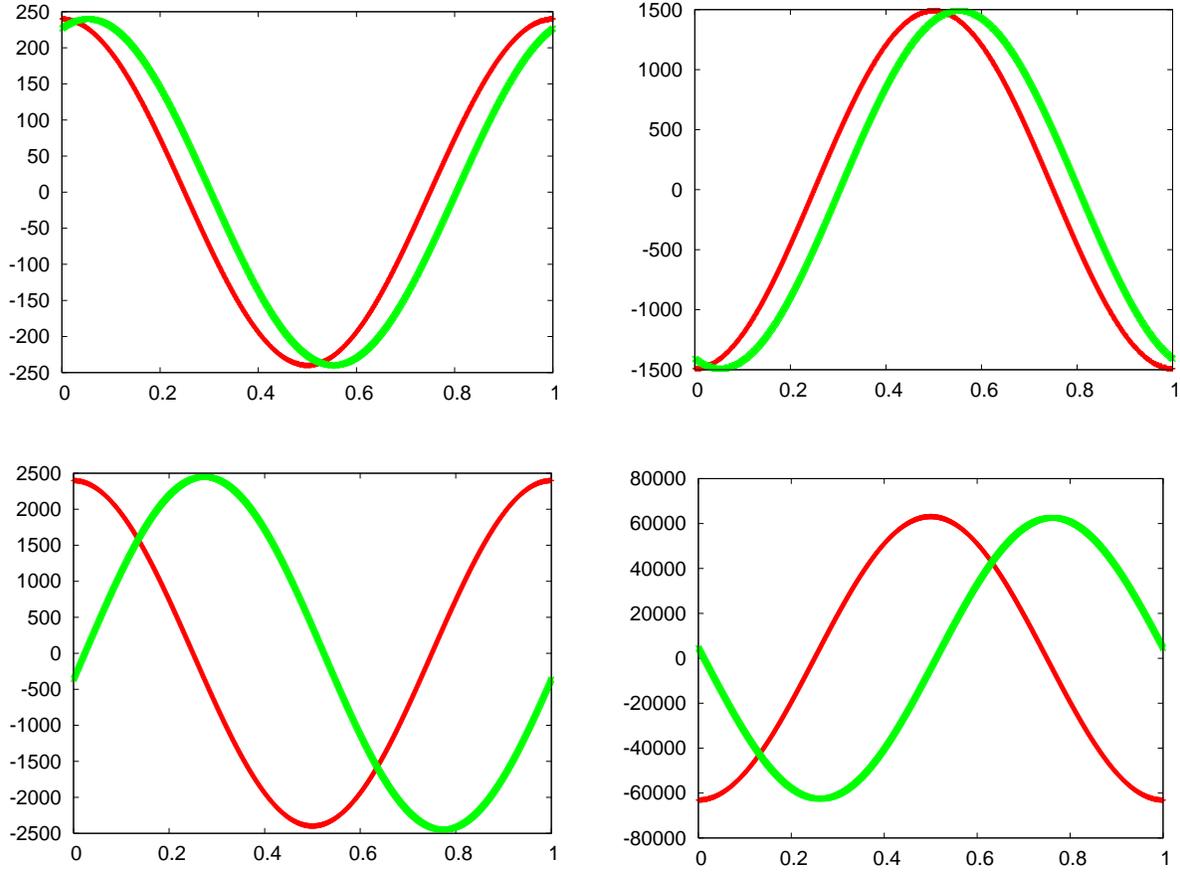}
\caption[\Alfven wave propagation in $z$ direction]
{Top panels:
Newtonian \Alfven wave propagation in $z$ direction.
The left figure show initial and final profile of $v_z$ and 
the right one does that of $B_z$.
Bottom panels: 
Relativistic \Alfven wave propagation in $z$ direction.
The left figure show initial and final profile of $v_z$ and 
the right one does that of $B_z$.}\label{relalz}
 \end{center}
\end{figure}
\clearpage

\section{Conservative Variable and Fundamental Variables}\label{CVFV}

In our numerical code, variables such as {$hWv^i,D,E,B^i$} are evolved.
These variables are called conservative variables.
It is necessary to compute 
fundamental variables such as $v^i,\rho ,e,B^i$ from these conservative variables.
In our computations, pressure is 
not described as an analytic function of energy $e$,
therefore algorithms used in other GRMHD simulations \citep{devi03} is not
available here. If the value of the enthalpy, $h$, is found,
the Lorentz factor, $W$, is obtained from the values of $hWv^{i}$
and then all the fundamental variables are obtained.
To determine the value of $h$, we search the root of the equation below:
\begin{eqnarray}
f(h)\stackrel{\mathrm{def}}{\equiv}(1+e/\rho+
p/\rho+\left|b\right|^2/\rho)-h=0 
\label{eq:ite}
\end{eqnarray}
where $\left|b\right|^2=(B^2+b_0^2)/W^2$.
The fundamental variables in the equation are obtained when $h$ is
assumed, 
as stated above.
We use simple bisection method to search the root of Eq. \ref{eq:ite}
and make the error of the equation, $\Delta h/h$, below $10^{-4}$.

\section{Test Problems}\label{TP}
At last we show some results of the numerical tests on our code.
We have done three typical problems
in both non-relativistic case and relativistic case.
Three problems are the shock tube problem, reflection shock problem
and magnetic shock tube problem.
We present the results one by one.

\paragraph{Shock Tubes}
We consider the shock tube
used by Sod \citep{sod78} in his comparison of finite difference scheme.
First we perform weak shock problem and strong shock problem in
Newtonian case \citep{hawl84b}. 
For the weak one,
the initial conditions of this problem
gas with $\Gamma=1.4$ with pressure and  density 
$P_{l}=1.0,\rho_{l}=10^5$ to the left of $x=0.5$
and 
$P_{r}=0.1,\rho_{r}=0.125\times 10^5$ to the right.
For the strong one,
the initial conditions of this problem are gas with $\Gamma=1.4$ with pressure and  density 
$P_{l}=0.67,\rho_{l}=100$ to the left of $x=0.5$
and 
$P_{r}=0.67\times 10^{-7},\rho_{r}=1.0$ to the right.
The numerical values greatly correspond to the analytic value.

Next we perform relativistic shock tube problems.
We fix the hydrodynamical variables
except for the pressure of the left side,
$\rho_{l}=10,\rho_{r}=1,p_{r}=10^{-6}$\citep{anni03}.
We set three types of pressure, i.e. 
$p_{l}=1.33$($W=1.08$) for lowly relativistic case,  
$p_{l}=6.67$($W=1.28$) fir mildly relativistic case 
and  $p_{l}=666.7$($W=3.28$) for highly relativistic case.
The results are shown in the left panels of Figure \ref{fig:rst}.
For lowly and mildly relativistic case,
the numerical value greatly correspond to the analytic value.
For highly relativistic case, 
the velocity of the numerical comparison 
doesn't reach that of the analytic solution.
It is due to the artificial viscosity which 
converted kinetic energy to internal energy.

\paragraph{Wall Reflections}
A second test presented here 
is the wall shock problem involving 
the shock heating of cold fluid hitting a wall at the
left boundary. When the fluid hits the wall
a shock forms and travels to the right, separating the pre-shocked
state composed of the initial data and the post-shocked state
with solution in the wall frame
\begin{eqnarray}
V_S      & = & \frac{\rho_{1} W_{1} V_{1}}{\rho_{2} - \rho_{1}W_{1}} , \\
P_{2}    & = & \rho_{2} (\Gamma - 1)(W_{1} - 1) , \\
\rho_{2} & = & \rho_{1} \left[ \frac{\Gamma + 1}{\Gamma - 1} +
               \frac{\Gamma}{\Gamma - 1}(W_{1} - 1) \right] , 
\end{eqnarray}
where $V_S$ is the velocity of the shock front, and the pre-shocked
energy and post-shocked velocity were both assumed
negligible ($e = V_2 = 0$).

The initial data are set up to be
uniform across the grid with adiabatic index $\Gamma=4/3$, 
pre-shocked density $\rho_{1} = 1$,
and pre-shocked pressure $P_{1} =  10^{-6}$.
And we change the velocity of the unshocked region parametrically.
The result is shown in the right panels of Figure \ref{fig:rst}.
For all computations, the differences 
between the numerical solution and the analytic one are small,
however in case of relativistic one,
pressure of the numerical solution is bigger than
the analytic one.
It is also due to the artificial viscosity assumed here.

\begin{figure}[ht]
 \begin{center}
  \includegraphics[width=.99\linewidth]{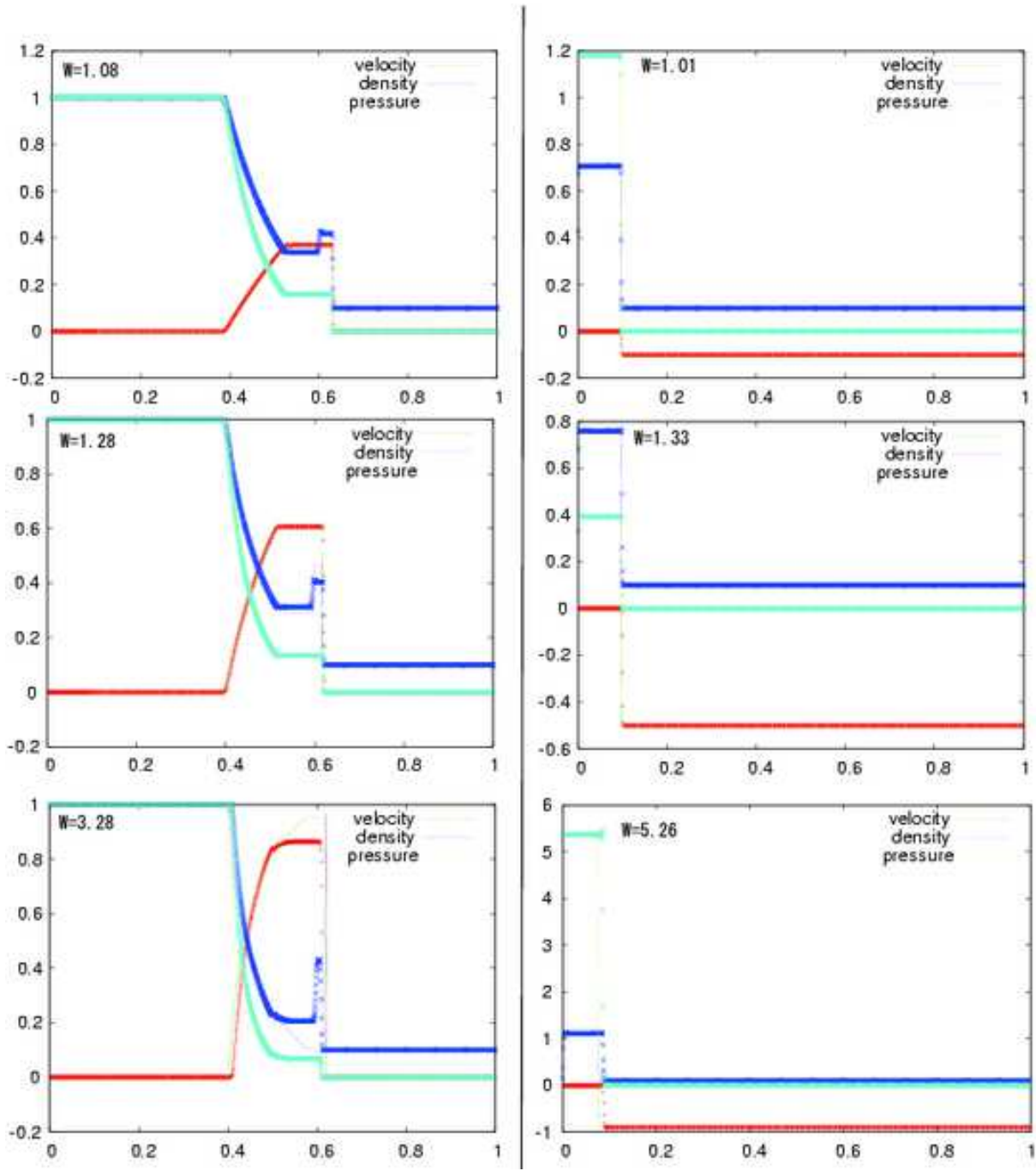}
\caption[Shock Tubes and Wall Reflection]{
Left and right panel show the shock tube and the wall reflection tests, 
respectively.
From top to bottom, the Lorentz factor becomes larger 
 (for right panels, top: $v=0.1c$ ($W=1.01$),
middle: $v=0.5c$($W=1.33$), and bottom:$v=0.9c$($W=5.26$)).
Note that the pressure and density is normalized by $100$.
}\label{fig:rst}
 \end{center}
\end{figure}
\clearpage

\paragraph{Magnetic Shock Tubes}
At last we present magnetic shock tube problems \citep{brio88}.
We show the initial condition and the results of the computations in
Table \ref{tab:bw}.
And we present mildly relativistic case in Figure \ref{fig:mrmst}, showing 
 that our code can handle the various magneto-sonic waves as good as
the code by \citet{devi03}.

\begin{figure}[ht]
 \begin{center}
  \includegraphics[width=.99\linewidth]{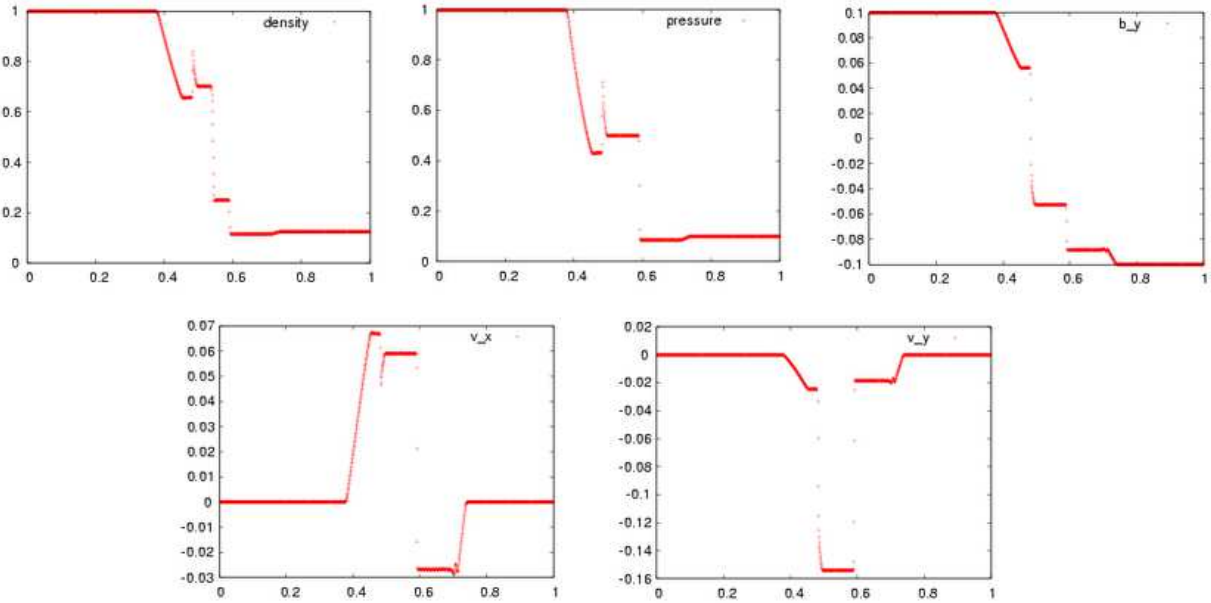}
\caption{Mildly relativistic magnetic shock tubes}\label{fig:mrmst}
 \end{center}
\end{figure}

\begin{table}[ht]
\caption[Initial and intermediate states for 
shock tube tests.]{Initial and intermediate states for 
shock tube tests. This table list the measured values
in each state of shock tube: ``Left'' is the initial left 
state for the given variables, 
``FR'' is the value of the variable
at the foot of leftmost fast rarefaction fan,
``SC'' is the value of the peak
of the slow compound wave,
``$\mathrm{CD_l}$'' is the value of left
of the constant discontinuity,
``$\mathrm{CD_r}$'' is the value of right
of the constant discontinuity,
``FR'' is the value of the variable
at the foot of the second fast rarefaction fan and 
``Right'' is the initial right 
state for the given variables, 
 }\label{tab:bw}

\begin{tabular}{llrrrrrrr}
 & & & & & & & & \\
\hline
Test & Variable & Left & FR & SC & $\mathrm{CD_l}$ & $\mathrm{CD_r}$ & FR & Right \\
\hline
\hline
Newtonian  & $\rho$& 1.00 & 0.66 & 0.84 & 0.70 & 0.25 & 0.12 & 0.13 \\
$B^x(\times 10^{-2})$     & $P (\times {10}^{-4})$   
                     & 1.00 & 0.44 & 0.73 & 0.50 & 0.50 & 0.09 & 0.10 \\
$ =0.75$ & $v^x (\times {10}^{-3})$   
                    & 0.00 & 6.67 & 4.6  & 5.99  & 6.02  & -2.79 & 0.00 \\
            & $v^y (\times {10}^{-2})$   
                    & 0.00 &-0.25 &-1.10 &-1.58 &-1.58 &-0.20 & 0.00 \\
    &  $\frac{B^y}{\sqrt{4\pi}} (\times {10}^{-2})$   
                    & 1.00 &0.6 &-0.5 &-0.03 &-0.54 &-0.9 &-1.00 \\
\hline
Mildly  & $\rho$& 1.00 & 0.65 & 0.84 & 0.70 & 0.24 & 0.11 & 0.13 \\
 Relativistic & $P (\times {10}^{-2})$   
& 1.00 & 0.42 & 0.71 & 0.49 & 0.49 & 0.08 & 0.10 \\
   $B^x(\times 10^{-1})$        & $v^x(\times {10}^{-1})$ & 0.00 & 0.67 & 0.46 & 0.58 & 0.58 &-0.26 & 0.00 \\
      $=0.75$   & $v^y(\times {10}^{-1})$
 & 0.00 &-0.24 &-0.94 &-1.5 &-1.5 &-1.9 & 0.00 \\
 & $\frac{B^y}{\sqrt{4\pi}}(\times {10}^{-1})$& 1.00 &0.56 &0.3 &-0.52 &-0.52 &-0.88 &-1.00 \\
\hline
Relativistic  & $\rho$& 1.00 & 0.59 & 0.70 & 0.65 & 0.31 & 0.11 & 0.13 \\
 $B^x$      & $P$   & 1.00 & 0.51 & 0.60 & 0.47 & 0.47 & 0.08 & 0.10 \\
      $=0.75$       & $v^x$ & 0.00 & 0.41 & 0.27 & 0.28 & 0.28 &-0.12 & 0.00 \\
            & $v^y$ & 0.00 &-0.07 &-0.62 &-0.58 &-0.11 &-0.11 & 0.00 \\
 & $\frac{B^y}{\sqrt{4\pi}}$& 1.00 &0.61 &0.19 &-0.24 &-0.24 &-0.81 &-1.00 \\
\hline
\hline
\end{tabular}
\end{table}

\end{document}